\newif\ifstocanon
\stocanonfalse
\documentclass[11pt]{article}

\usepackage[utf8]{inputenc}
\usepackage[T1]{fontenc}
\usepackage{lmodern}
\usepackage[protrusion=true,expansion=true]{microtype}
\usepackage[margin=1in]{geometry}
\usepackage{amsmath,amssymb,amsthm,mathtools}
\usepackage{booktabs,array,tabularx}
\usepackage{enumitem}
\usepackage{float,caption}
\usepackage[dvipsnames,table]{xcolor}
\usepackage{fancyhdr}
\usepackage{tikz}
\usetikzlibrary{arrows.meta,positioning,calc,decorations.pathreplacing}

\definecolor{mahogany}{rgb}{0.70, 0.20, 0.05}
\definecolor{darkblue}{rgb}{0.05, 0.20, 0.60}
\definecolor{commentpurple}{rgb}{0.48, 0.14, 0.52}

\usepackage[colorlinks=true,linkcolor=darkblue,citecolor=mahogany,urlcolor=darkblue]{hyperref}
\usepackage[capitalize,nameinlink]{cleveref}

\allowdisplaybreaks
\fancypagestyle{pg}{
  \fancyhf{}
  \cfoot{--\ \thepage\ --}
  
}

\newtheorem{theorem}{Theorem}[section]
\newtheorem{lemma}[theorem]{Lemma}
\newtheorem{proposition}[theorem]{Proposition}

\newcommand{\innercustomthm}{}
\newtheorem*{theorem*}{Theorem~\innercustomthm}
\newtheorem*{lemma*}{Lemma~\innercustomthm}
\newtheorem*{corollary*}{Corollary~\innercustomthm}
\theoremstyle{definition}
\newtheorem{definition}[theorem]{Definition}
\newtheorem{remark}[theorem]{Remark}

\crefname{theorem}{Theorem}{Theorems}
\Crefname{theorem}{Theorem}{Theorems}
\crefname{lemma}{Lemma}{Lemmas}
\Crefname{lemma}{Lemma}{Lemmas}
\crefname{proposition}{Proposition}{Propositions}
\Crefname{proposition}{Proposition}{Propositions}
\crefname{corollary}{Corollary}{Corollaries}
\Crefname{corollary}{Corollary}{Corollaries}
\crefname{claim}{Claim}{Claims}
\Crefname{claim}{Claim}{Claims}
\crefname{definition}{Definition}{Definitions}
\Crefname{definition}{Definition}{Definitions}
\crefname{remark}{Remark}{Remarks}
\Crefname{remark}{Remark}{Remarks}
\crefname{section}{Section}{Sections}
\Crefname{section}{Section}{Sections}
\crefname{appendix}{Appendix}{Appendices}
\Crefname{appendix}{Appendix}{Appendices}
\crefname{table}{Table}{Tables}
\Crefname{table}{Table}{Tables}
\crefname{figure}{Figure}{Figures}
\Crefname{figure}{Figure}{Figures}

\floatstyle{ruled}
\newfloat{algorithm}{tbp}{loa}
\floatname{algorithm}{Algorithm}
\crefname{algorithm}{Algorithm}{Algorithms}
\Crefname{algorithm}{Algorithm}{Algorithms}

\newlist{algsteps}{enumerate}{3}
\setlist[algsteps]{label=\arabic*:, ref=\arabic*, leftmargin=2.2em, itemsep=2.5pt, topsep=4pt, parsep=0pt, before=\raggedright}
\crefname{algstepsi}{Line}{Lines}
\Crefname{algstepsi}{Line}{Lines}

\newcommand{\algcomment}[1]{\hfill $\triangleright$~{\small\color{commentpurple} #1}}

\makeatletter
\newcounter{restate@savedeq}
\NewDocumentEnvironment{restatable}{O{} m m +b}{%
  \expandafter\xdef\csname #3@eq\endcsname{\the\c@equation}%
  \expandafter\gdef\csname #3\endcsname{%
    \begingroup
    \setcounter{restate@savedeq}{\value{equation}}%
    \setcounter{equation}{\csname #3@eq\endcsname}%
    \renewcommand{\theHequation}{restate.#3.\theequation}%
    \renewcommand{\label}[1]{}%
    \let\ltx@label\@gobble
    \let\label@in@display\@gobble
    \ifx&#1&%
      \begin{#2*}#4\end{#2*}%
    \else%
      \begin{#2*}[#1]#4\end{#2*}%
    \fi%
    \setcounter{equation}{\value{restate@savedeq}}%
    \endgroup
  }%
  \ifx&#1&%
    \begin{#2}#4\end{#2}%
  \else%
    \begin{#2}[#1]#4\end{#2}%
  \fi%
}{}
\newcommand{\restatethm}[2]{%
  \begingroup
  \renewcommand{\innercustomthm}{\ref{#2}}%
  \csname #1\endcsname
  \endgroup
}
\makeatother

\newcommand{\E}{\mathbb{E}}
\newcommand{\R}{\mathbb{R}}
\newcommand{\N}{\mathbb{N}}

\newcommand{\supp}{\operatorname{supp}}
\newcommand{\nnz}{\operatorname{nnz}}
\newcommand{\poly}{\operatorname{poly}}

\newcommand{\norm}[1]{\left\lVert #1 \right\rVert}
\newcommand{\abs}[1]{\left\lvert #1 \right\rvert}

\newcommand{\dens}[1]{\E[#1]}
\newcommand{\pot}{C_{1/4,1/4}}

\newcommand{\Prune}{\textnormal{\textsc{Prune}}}
\newcommand{\EstimateNorm}{\textnormal{\textsc{EstimateNorm}}}
\newcommand{\Sift}{\textnormal{\textsc{Sift}}}
\newcommand{\FindRect}{\textnormal{\textsc{FindRect}}}
\newcommand{\DecomposeTwoD}{\textnormal{\textsc{Decompose2D}}}
\newcommand{\FindCube}{\textnormal{\textsc{FindCube}}}
\newcommand{\DecomposeThreeD}{\textnormal{\textsc{Decompose3D}}}

\newcommand{\tagreg}{\textnormal{\textsc{reg}}}
\newcommand{\tagdenseA}{\textnormal{\textsc{denseA}}}
\newcommand{\tagdenseB}{\textnormal{\textsc{denseB}}}
\newcommand{\tagnormal}{\textnormal{\textsc{normal}}}
\newcommand{\tagsparse}{\textnormal{\textsc{sparse}}}

\begin{document}

\title{Combinatorial Boolean Matrix Multiplication in \texorpdfstring{$n^3 / 2^{\Omega(\sqrt{\log n})}$}{n\^{}3 / 2\^{}Omega(sqrt(log n))} Time}
\ifstocanon
\author{Anonymous Author(s)}
\else
\author{
Piaoyang Cui \\ Google \and
Yiming Liu \\ Google  \and
David P.~Woodruff \\ Carnegie Mellon University \& Google 
}
\fi
\date{}
\hypersetup{pageanchor=false}
\setcounter{page}{0}
\maketitle
\thispagestyle{empty}

\begin{abstract}
For over fifty years, combinatorial algorithms for Boolean matrix multiplication (BMM) and triangle detection saved
only polylogarithmic factors over cubic time. At STOC 2024, Abboud, Fischer, Kelley, Lovett, and Meka broke this
barrier with a deterministic $O(n^3 / 2^{\Omega((\log n)^{1/7})})$-time algorithm, and asked whether
$n^3 / 2^{\Omega(\sqrt{\log n})}$ time is achievable. We answer this question in the affirmative, giving deterministic
combinatorial algorithms for both BMM and triangle detection running in $O(n^3 / 2^{\Omega(\sqrt{\log n})})$ time. 

Following Abboud et al., we decompose $(A,B)$ into pairs of submatrices $(A_k,B_k)$, called \emph{pieces}, each of
which is either sparse or pseudorandom in a sense that forces $A_k\star B_k$ to be nearly all-ones on its vertex sets.
Whereas their decomposition produces $2^{O(d^7)}$ pieces at regularity parameter $d$, we obtain $2^{O(d^2)}$ pieces,
directly improving the runtime exponent from $(\log n)^{1/7}$ to $\sqrt{\log n}$. Prior work found pseudorandom pieces
by repeatedly restricting to denser submatrices, paying worst-case vertex losses at each step that compounded across
nested recursion levels, together with an expensive derandomization step. We eliminate the extra powers of $d$ by
matching the vertex loss at each step to the size of its density jump so losses telescope across scales, derandomizing
witness search via $L_d$-norm estimation so the sampler length stays $2^{O(d)}$, and bounding overlap across the
recursion with a fractional density-weighted volume budget rather than raw submatrix volume.
\end{abstract}

\clearpage
\hypersetup{pageanchor=true}
\pagestyle{pg}
\setcounter{page}{1}

\section{Introduction}\label{sec:intro}

\subsection{Boolean Matrix Multiplication and Combinatorial Algorithms}\label{sec:history}

Given binary $n\times n$ matrices $A,B$, their Boolean product $A\star B$ records whether each pair $(x,z)$ has
a witness $y$ with $A(x,y)=B(y,z)=1$. Algebraic algorithms use signed ring identities to compute this product in
$O(n^{\omega+o(1)})$ time, where $\omega$ is the matrix multiplication exponent~\cite{Strassen69,DWZ23,VXXZ24}. Such cancellations do not extend
directly to many related graph problems~\cite{VassilevskaWY07,WilliamsW18,AbboudFS24}, motivating
\emph{combinatorial} algorithms that work directly with the graph structure~\cite{BansalW12,WilliamsW18,AFKLM24}.
BMM is subcubically equivalent to triangle detection by combinatorial reductions~\cite{WilliamsW18}.
We use the tripartite form: $A$ gives edges on $X\times Y$, $B$ on $Y\times Z$, and $C$ on $X\times Z$.
A triangle is a $2$-path $x\to y\to z$ through $A,B$ plus its closing edge in $C$.

\begin{table}[htb]
\centering
\footnotesize
\setlength{\tabcolsep}{3pt}
\renewcommand{\arraystretch}{1.08}
\begin{tabularx}{\textwidth}{@{}l l >{\raggedright\arraybackslash}X@{}}
\toprule
\textbf{Reference} & \textbf{Running time} & \textbf{Main technique} \\
\midrule
Arlazarov et al.~\cite{ADKF70} & $O(n^3/\log^2 n)$ & Four-Russians lookup tables \\
Bansal--Williams~\cite{BansalW12} & $O(n^3(\log\log n)^2/(\log n)^{9/4})$ & Weak regularity and packing \\
Chan~\cite{Chan15} & $O(n^3(\log\log n)^3/\log^3 n)$ & Word-level divide-and-conquer \\
Yu~\cite{Yu18} & $\hat O(n^3/\log^4 n)$ & Compressed tabulation \\
Abboud et al.~\cite{AFKLM24} & $O(n^3/2^{\Omega((\log n)^{1/7})})$ & Grid regularity; no packing \\
\textbf{This paper} & $\boldsymbol{O(n^3/2^{\Omega(\sqrt{\log n})})}$ & \textbf{Dyadic grid regularity; no packing} \\
\bottomrule
\end{tabularx}
\caption{Progress on combinatorial Boolean matrix multiplication and triangle detection. The first four bounds
use bit packing; $\hat O$ hides $(\log\log n)^{O(1)}$ factors.}
\label{tab:bmm-history}
\end{table}

At STOC 2024, Abboud, Fischer, Kelley, Lovett, and Meka (AFKLM)~\cite{AFKLM24} broke the polylogarithmic barrier
(\Cref{tab:bmm-history}). Building on~\cite{KelleyM23,KelleyLM23}, they used a pseudorandomness notion called
\emph{grid regularity}. The \emph{density} of a binary matrix is the fraction of its entries that are ones. For an
integer parameter $d$ and small constant $\varepsilon>0$, a binary matrix of density $p$ is
\emph{$(\varepsilon,2,d)$-regular} (or simply \emph{regular}) if, for two random rows and $d$ random
columns chosen independently with replacement, all $2d$ selected entries are ones with probability at most
$((1+\varepsilon)p)^{2d}$. By convexity, every matrix of density $p$ has probability at least $p^{2d}$, and a
random matrix nearly attains this minimum. Intuitively, a regular matrix has no hidden dense spots. If some set
of rows pairwise share far more columns than random rows do, then grids inside that set inflate the probability,
and the $d$-th power makes this visible even when the set is exponentially small in $d$.

The Kelley--Lovett--Meka (KLM) theorem says that if $A$ and $B^T$ (the transpose of $B$) are regular and no row of $A$
or column of $B$ has degree below $(1-\varepsilon)$ times the average, then, as in a random graph, almost every pair
$(x,z)\in X\times Z$ has a witness: $A\star B$ is all-ones outside a $2^{-\varepsilon d/2}$ fraction of entries.
AFKLM decomposed $(A,B)$ into $2^{O(d^7)}$ \emph{pieces}, pairs of submatrices $(A_k,B_k)$ on shared middle vertex sets
that are either \emph{regular} (both $A_k$ and $B_k^T$ are regular with no low-degree rows) or \emph{sparse} (at least
one of $A_k,B_k$ has density at most $2^{-d}$, or $B_k$ has fewer than $2^d$ entries; see \Cref{def:pieces}). They asked
whether the saving of this approach could reach $2^{\Omega(\sqrt{\log n})}$~\cite[Section~3.4]{AFKLM24}. We reduce the
piece bound to $2^{O(d^2)}$ and answer this question (see below).
Our saving is still $n^{o(1)}$, consistent with the Combinatorial BMM hypothesis, which conjectures that no
$O(n^{3-\eta})$-time combinatorial BMM algorithm exists for any constant
$\eta>0$~\cite{AbboudVW14,WilliamsW18,AFKLM24}.

\subsection{Our Results}\label{sec:results}

\begin{restatable}[Combinatorial BMM and triangle detection]{theorem}{thmmainbmm}\label{thm:main-bmm}
There is a deterministic combinatorial algorithm that multiplies two $n \times n$ Boolean matrices, and one that
decides whether an $n$-vertex graph contains a triangle, in time
\begin{equation}\label{eq:main-bmm-time}
    O\!\left(\frac{n^3}{2^{\Omega(\sqrt{\log n})}}\right).
\end{equation}
\end{restatable}
\Cref{thm:main-bmm} has been machine-checked in Lean~4; the formalization follows the full
version of this paper, and both are available in the
\href{https://github.com/lockqueue/bmm}{accompanying repository} (see \Cref{sec:ai-disclosure}).

We note that we use the $O(\log n)$-bit Word RAM with comparisons, array indexing, and elementary integer arithmetic
($+,-,\times,\div$), but no bitwise operations, bit packing, floating-point arithmetic, or Four-Russians tables.
Known reductions give corresponding improvements for $k$-Clique and $k$-Dominating Set (for fixed $k\ge3$),
All-Edges Triangle, context-free parsing, transitive closure, and Boolean binary join-project queries
(\Cref{app:applications}).

For triangle detection, the algorithm has two phases: a \emph{preprocessing phase} that decomposes $(A,B)$ into
$K=2^{O(d^c)}$ pieces without looking at $C$, and a \emph{query phase} that tests each piece against $C$.
In general, producing $K=2^{O(d^c)}$ pieces takes $O(n^{5/2}2^{O(d^c)})$ preprocessing time and
$O(n^2 2^{O(d^c)}+n^3/2^{\Omega(d)})$ total query time across all pieces.
Choosing $d=\Theta((\log n)^{1/c})$ with a sufficiently small leading constant keeps preprocessing at
$O(n^{11/4})$, which is negligible. AFKLM had $c=7$, while we obtain $c=2$.

To state our decomposition more precisely, we write $\dens{M}$ for the fraction of ones in a binary matrix $M$. A matrix is \emph{$\varepsilon$-min-degree} if
every row has at least $(1-\varepsilon)$ times the average row degree. 

\begin{definition}[Pieces, sparsity, and regularity]\label{def:pieces}
Fix $\varepsilon := 1/160$, let $d \ge 640$ be an even integer, and let $X, Y, Z$ be finite non-empty sets. A
\emph{piece} is a tuple $(X_k, Y_k, Z_k, A_k, B_k)$, where $X_k \subseteq X$, $Y_k \subseteq Y$, and $Z_k \subseteq Z$
are non-empty subsets and $A_k \in \{0,1\}^{X_k \times Y_k}$ and $B_k \in \{0,1\}^{Y_k \times Z_k}$ are binary matrices.
We say that a piece is:
\begin{itemize}[leftmargin=*,nosep]
    \item \emph{sparse} if $\min(\dens{A_k}, \dens{B_k}) \le 2^{-d}$ or $\abs{Y_k}\abs{Z_k} < 2^d$; and
    \item \emph{regular} if $\min(\dens{A_k}, \dens{B_k}) > 2^{-d}$ and both $A_k$ and $B_k^T$ are
    $(\varepsilon, 2, d)$-regular (informally defined above, formally defined in \Cref{def:grid-norm}) and $\varepsilon$-min-degree.
\end{itemize}
\end{definition}

Our decomposition is as follows: 
\begin{restatable}[Three-dimensional regularity decomposition]{theorem}{thmmaindecomp}\label{thm:main-decomp}
Fix $\varepsilon := 1/160$ and let $d \ge 640$ be an even integer. There is a deterministic algorithm that, given finite
non-empty sets $X, Y, Z$ with $n := \abs{X} + \abs{Y} + \abs{Z}$ and binary matrices $A \in \{0, 1\}^{X \times Y}$ and
$B \in \{0, 1\}^{Y \times Z}$, computes in time $O(n^{5/2} 2^{O(d^2)})$ a list
$\mathcal{S} = \{(X_k, Y_k, Z_k, A_k, B_k)\}_{k=1}^K$ of $K \le 2^{O(d^2)}$ pieces with $A_k \le A[X_k, Y_k]$ and
$B_k \le B[Y_k, Z_k]$ entrywise (where $A[X_k,Y_k]$ denotes the submatrix of $A$ restricted to $X_k\times Y_k$) such that:
\begin{enumerate}[leftmargin=*,nosep]
    \item $AB = \sum_{k=1}^K A_k B_k$ as an exact identity of integer matrix products, where each $A_k B_k$ is extended
    by zeros outside $X_k \times Z_k$;
    \item every piece in $\mathcal{S}$ is either sparse or regular; and
    \item the fractional density-weighted potential satisfies
    \begin{equation}\label{eq:main-potential}
    \begin{aligned}
        \pot(\mathcal{S}) &:= \sum_{k=1}^K \dens{A_k}^{1/4}\dens{B_k}^{1/4}\abs{X_k}\abs{Y_k}\abs{Z_k} \\
        &\le 1024\,\dens{B}^{1/4}\abs{X}\abs{Y}\abs{Z} \le 1024\,\abs{X}\abs{Y}\abs{Z}.
    \end{aligned}
    \end{equation}
\end{enumerate}
\end{restatable}

\paragraph{How to read the decomposition theorem:}
The construction proves Clause~1 by partitioning \emph{$2$-paths}, not vertices or edges.
Clauses~2 and~3 make the pieces easy to test for triangles and bound the total cost despite potentially overlapping vertex sets in the partition. 

\section{Overview of the Algorithm and Main Ideas}\label{sec:overview}

In this overview, $n:=\abs{X}+\abs{Y}+\abs{Z}$, $\varepsilon=1/160$, and $d\ge640$ is even.
Write $q:=2^{-\varepsilon d/2}=2^{-d/320}$ for KLM's missing-pair bound.
We first explain how to use the pieces, then how to find them.

\subsection{What a Piece Is, and What Testing It Against \texorpdfstring{$C$}{C} Means}\label{sec:overview-query}

\paragraph{The role of $C$.}
Triangle detection asks whether an edge $(x,z)$ of $C$ is closed by a $2$-path through $A,B$.
To test piece $k$, we restrict the third graph to $C_k:=C[X_k,Z_k]$ and ask whether
$(A_k,B_k,C_k)$ has a triangle. A triangle found there is an original triangle, since the piece contains only
original edges. Conversely, an original triangle's $2$-path belongs to a piece, by the routing argument in
\Cref{sec:overview-search} below, and its closing edge belongs to that piece's $C_k$.

We never decompose $C$: the restrictions $C_k$ may overlap, so a closing edge can be tested several times.
Plain BMM has no third input; the reduction of \Cref{prop:bmm-to-tridet} supplies matrices $C$ marking the
output entries being asked about.

\paragraph{Binary pieces, integer products, and no cancellations.}
Every operation our decomposition performs on $A$ and $B$ only restricts to vertex subsets or deletes edges, so all
piece matrices $A_k,B_k$ remain binary ($0/1$) subgraphs of $A[X_k,Y_k]$ and $B[Y_k,Z_k]$ (not necessarily induced,
because earlier steps may already have deleted some of their edges).
In \Cref{thm:main-decomp}, the product $AB$ is taken over the integers: $(AB)(x,z)$ counts witnesses $y\in Y$, and the
Boolean product $A\star B$ is its support (the set of pairs with $(AB)(x,z)\ge 1$).
Unlike a Strassen-style signed identity, these nonnegative witness counts cannot cancel across pieces, so $(AB)(x,z)>0$
if and only if $(A_k B_k)(x,z)>0$ for at least one piece $k$, allowing us to test pieces independently.

\paragraph{What regularity gives us.}
Equivalently to the definition in \Cref{sec:history}, the \emph{grid norm} $\norm{A}_{U(2,d)}$ is the $2d$-th root of
the probability that two random rows and $d$ random columns form an all-ones $2\times d$ grid, so $A$ is
$(\varepsilon,2,d)$-regular when $\norm{A}_{U(2,d)}\le(1+\varepsilon)\dens{A}$.
Write $u:=\dens{A_k}$ and $v:=\dens{B_k}$. For a regular piece, KLM (\Cref{prop:regular-product}) puts its
normalized $2$-path count $(A_k\circ B_k)(x,z):=(A_kB_k)(x,z)/\abs{Y_k}$ between $uv/2$ and $3uv/2$ on all but a
$q$-fraction of $X_k\times Z_k$. In particular, at most a $q$-fraction has no $2$-path; the minimum-degree condition is
needed because nearly isolated vertices in $X_k$ or $Z_k$ could otherwise produce many empty pairs.

\paragraph{Fast ways to answer.}
Let $V_k:=\abs{X_k}\abs{Y_k}\abs{Z_k}$ be the piece's volume. We first count the edges of
$A_k,B_k,C_k$ and build their supports. This costs $O(n^2)$ per piece, including reading $C_k$, and gives us an 
$n^2K$ term in the query time.
\begin{enumerate}[leftmargin=*,nosep,topsep=2pt,itemsep=2pt]
    \item \emph{The middle-right grid is tiny: $\abs{Y_k}\abs{Z_k}<2^d$.}
    We try every triple. The $X$-side may still be large, but there are fewer than $n2^d$ triples:
    $O(n^{3/2})$ per piece and $O(n^{7/4})$ altogether at our final parameters which we set below.
    This handles granularity: even one $B$-edge on such a grid already has density greater than $2^{-d}$.
    \item \emph{Some graph has density at most $q$.}
    Scan the edges of the sparsest of $A_k,B_k,C_k$. An edge fixes two corners; we try the third and look up the
    other two edges. A $C_k$-edge leaves $y\in Y_k$ to try, an $A_k$-edge leaves $z\in Z_k$, and a $B_k$-edge
    leaves $x\in X_k$. Every triangle uses one scanned edge, so this is exhaustive.
    The cost is the smallest density times $V_k$.
    \item \emph{All three graphs have density greater than $q$.}
    The piece is not tiny, and $u,v>q>2^{-d}$, so it is not sparse and must be regular.
    KLM leaves at most $q\abs{X_k}\abs{Z_k}$ pairs without a $2$-path, whereas $C_k$ has more edges than this.
    Its edges cannot all hide among the missing pairs. In this case we can answer YES by the pigeonhole principle, without finding a witness $y$.
\end{enumerate}
Using KLM's threshold $q$ leaves no gap; if every piece fails, we answer NO.

For illustration, take a regular piece with $m$ vertices per side and pretend $q=1/1000$.
At most $m^2/1000$ pairs lack a common neighbor. If $C_k$ has more edges, a triangle is guaranteed.
Otherwise, checking its edges against all $m$ middle vertices costs at most $m^3/1000$.

\begin{remark} 
We note that zero-padding every piece to the full vertex sets preserves the above integer identity, but destroys some of the properties that we claim. Complete blocks occupying $1/100$ of each side, for example, have a product supported on only a 
$10^{-4}$ fraction of $X\times Z$. A graph $C$ can be dense globally and miss that rectangle entirely.
So when we apply the pigeonhole principle above, it must be applied inside $X_k\times Z_k$. 
\end{remark}

\paragraph{Why a per-piece saving is not enough.} A scan costs at most $qn^3$, but the sum $Kqn^3$ need not be any better than $O(n^3)$ time: the bound
$K=2^{O(d^2)}$ will overwhelm the savings  $q=2^{-\Theta(d)}$.
Nor can we simply sum $qV_k$, since the pieces are not necessarily disjoint. 

We instead charge Case~2 to the \emph{weighted volume} $\pot(\mathcal{S})$ in \Cref{eq:main-potential}.
Put $m_k:=\min(u,v,\dens{C_k})$ and $t:=\min(u,v)\le\sqrt{uv}$. In Case~2 ($m_k\le q$), scanning the sparsest graph
checks $m_k V_k$ triples, and in both regimes $t\le q$ and $t>q$ we have $m_k\le\sqrt{q}\,(uv)^{1/4}$:
\[
\begin{array}{ll}
t\le q: & m_k\le t = \sqrt{t}\,\sqrt{t} \le \sqrt q\,\sqrt t\le \sqrt q\,(uv)^{1/4},\\
t>q: & m_k\le q = \sqrt{q}\,\sqrt{q} < \sqrt q\,(uv)^{1/4} \quad\text{(since $u,v\ge t>q$)}.
\end{array}
\]
Since $q=2^{-d/320}$, we have $\sqrt{q}=2^{-d/640}$, so Case~2 checks at most
$\sqrt{q}\,(uv)^{1/4}V_k = 2^{-d/640}(uv)^{1/4}V_k$ triples on piece $k$.
Meanwhile, Case~1 ($\abs{Y_k}\abs{Z_k}<2^d$) checks at most $n2^d=O(n^{3/2})$ triples per piece at our eventual choice
$2^d=O(\sqrt n)$, which is absorbed into the $O(n^2K)$ setup overhead. Summing across all $K$ pieces gives total query time
$O(n^2K+2^{-d/640}\pot(\mathcal{S}))=O(n^2K+n^3/2^{d/640})$: we trade a per-piece factor of $q$ for $\sqrt q$, gaining a
weighted volume that can be summed across overlapping pieces.

\subsection{Finding Pieces by Zooming In, and Counting Them}\label{sec:overview-search}

We retain AFKLM's general framework of minimum-degree pruning, cube search, and the KLM theorem.
Its main idea is that a matrix is either regular with adequate minimum degree, or a suitable restriction
makes it denser. 

It will be important to try to find regular rectangles; for the moment let's ignore some other constraints that need to be considered when finding such rectangles. 

\paragraph{A zoom keeps above-average structure.}
To find a single regular rectangle inside a 2D matrix $A$ of density $\mu>2^{-d}$, our search subroutine \FindRect{}
alternates between two steps---minimum-degree pruning (\Prune{}) and regularity sifting (\Sift{})---measuring row degrees
as fractions of the current column set:
\begin{itemize}[leftmargin=*,nosep,topsep=2pt,itemsep=2pt]
    \item \emph{Pruning (\Prune{}, \Cref{lem:min-degree}):} Repeatedly delete any row whose degree is below
    $(1-\varepsilon)$ times the \emph{current} average degree, stopping either when all surviving rows satisfy the
    $\varepsilon$-min-degree condition or when at least half the rows have been deleted. Deleting below-average rows can
    only raise the average: if $R$ is the set of deleted rows and $\mu'$ is the final density, each deleted row had
    degree below $(1-\varepsilon)\mu'$, so if $\abs{R}\ge\abs{X}/2$, then
    \[
        \mu' > \frac{\mu}{1-\varepsilon\abs{R}/\abs{X}} \ge (1+\varepsilon/2)\mu.
    \]
    For instance, if half the rows are all-zero and the rest have degree $2\mu$, deleting the zero rows doubles the
    density (and a positive-density matrix can never lose all its ones). Whenever $\mu'\ge(1+\varepsilon/2)\mu$---even if
    the $\varepsilon$-min-degree condition already holds---we restart the search on the surviving rows.
    \item \emph{Sifting (\Sift{}, explained in \Cref{sec:overview-sift}):} Otherwise, the surviving rows are
    $\varepsilon$-min-degree with density $\mu'\ge\mu$. The routine \Sift{} tests whether the current submatrix is
    $(\varepsilon,2,d)$-regular. If it is, \FindRect{} returns this regular rectangle; if not, \Sift{} finds a column
    subset and a subset of rows whose degrees into those columns are at least $2^s(1+\varepsilon/2)\mu'$ for some integer
    scale $s\in\{0,\dots,d-1\}$, and restarts the search on that submatrix.
\end{itemize}
Every restart multiplies the density by at least $1+\varepsilon/2$. Since the  density starts above $2^{-d}$ and never
exceeds $1$, \FindRect{} performs at most $2d/\varepsilon=O(d)$ zooms before returning a regular rectangle.

\paragraph{Peeling a 2D matrix without losing pruned edges.}
Suppose we run \FindRect{} on $A$ and it returns a regular submatrix $A_1=A[X_1,Y_1]$. Along the way, \FindRect{} threw
away many rows and columns of $A$ to zoom in on $X_1\times Y_1$. We cannot simply discard the 1-entries of $A$ outside
$X_1\times Y_1$. Instead, our 2D decomposition routine \DecomposeTwoD{} records $A_1$ as the first output rectangle,
\emph{zeros out only the entries inside $X_1\times Y_1$} in $A$, and leaves every other entry of $A$ intact---including all
1-entries in rows and columns that \FindRect{} pruned during its zoom. It then runs \FindRect{} again from scratch on
the full vertex sets $X\times Y$ of this updated matrix, extracts a second rectangle $A_2$, zeros out $A_2$'s entries,
and repeats until the remaining matrix has density at most $2^{-d}$ (or, once we add density caps in
\Cref{sec:overview-depth}, falls below a density floor), at which point that remainder is output as well. Because each step zeros out
precisely the 1-entries it outputs, the output matrices have disjoint supports and sum to $A$.

Suppose each extracted rectangle $A_\ell=A^{(\ell)}[X_\ell,Y_\ell]$ occupies at least an $f$-fraction of the original
area $\abs{X}\abs{Y}$. Let $t_\ell:=\dens{A^{(\ell)}}$ be the remaining density before step $\ell$ (with $t_1\le 1$).
Because zooming never decreases density, $\dens{A_\ell}\ge t_\ell$, so zeroing out $A_\ell$ leaves density
\[
    t_{\ell+1}
    = t_\ell - \frac{\abs{X_\ell}\abs{Y_\ell}}{\abs{X}\abs{Y}}\dens{A_\ell}
    \le (1-f)t_\ell.
\]
After $k$ extractions, $t_{k+1}\le (1-f)^k t_1 \le e^{-fk}$, which is at most $2^{-d}$ once
$k\ge (d\ln 2)/f = O(d/f)$.
In \Cref{sec:overview-sift}, we bound the cumulative row and column losses inside \FindRect{} by $2^{-O(d^2)}$,
giving $f\ge2^{-O(d^2)}$ and at most $2^{O(d^2)}$ rectangles.

\paragraph{Aligning the two matrices in 3D.}
It is important to note that independent 2D decompositions of $A$ and $B$ do not suffice: in every regular piece $(A_k,B_k)$, both $A_k$ and $B_k^T$
must be regular and $\varepsilon$-min-degree on the \emph{same} middle vertex set $Y_k$.
Our 3D search routine \FindCube{} therefore searches for a block $B[Y^*,Z^*]$ of $B$ together with a compatible 2D
decomposition of $A[X,Y^*]$. Given a candidate block $B[Y^*,Z^*]$, it first prunes low-degree $Y$-rows of $B[Y^*,Z^*]$ so
that restricting to any subset $Y_\ell\subseteq Y^*$ cannot substantially lower $B$'s density, and then runs
\DecomposeTwoD{} on $A[X,Y^*]$ to partition it into rectangles $A_\ell$ on $X_\ell\times Y_\ell$ (plus a sparse or
low-density 2D remainder on $X\times Y^*$).

If for some $A_\ell$ the corresponding submatrix $B[Y_\ell,Z^*]$ is much denser than $B[Y^*,Z^*]$, \FindCube{} zooms in
to $B[Y_\ell,Z^*]$. Otherwise, for each regular rectangle $A_\ell$, we want $B^T[Z^*,Y_\ell]$ to be regular and
$\varepsilon$-min-degree as well, so we run \Prune{} and \Sift{} on the rows of $B^T[Z^*,Y_\ell]$ (which are vertices of
$Z^*$). If pruning deletes at least a $\gamma$-fraction of $Z^*$ (where say $\gamma:=2^{-12}=1/4096$), or raises density too
much, or \Sift{} finds a denser witness submatrix, we get a denser sub-block of $B$ and restart \FindCube{} on that new
$B$-block while keeping \emph{all} of $A$ restricted to its new middle set.
When every regular $A_\ell$ passes, the surviving set $Z_\ell\subseteq Z^*$ omits only a tiny set
$R_\ell:=Z^*\setminus Z_\ell$ of fewer than $\gamma\abs{Z^*}$ vertices. This yields aligned regular pieces
$(A_\ell,B[Y_\ell,Z_\ell])$, while the omitted $Z$-vertices $R_\ell$, the 2D remainder of $A[X,Y^*]$, any rectangles where
$A$ or $B$ jumped above a constant-factor density cap (which we defer as \emph{boosted} children; see
\Cref{sec:overview-depth}), and the leftover $B$-edges $B-B[Y^*,Z^*]$ are handled recursively by our main 3D
procedure \DecomposeThreeD{}.

\paragraph{Following one $2$-path proves exactness.}
Although vertex sets and even individual edges can appear in multiple recursive calls, every $2$-path $(x,y,z)$ with
$A(x,y)=B(y,z)=1$ is routed to a \emph{unique} child or output piece of \DecomposeThreeD{}:
\begin{enumerate}[leftmargin=*,nosep,topsep=2pt,itemsep=2pt]
    \item \emph{Outside vs.\ inside the extracted $B$-block:} If $(y,z)\notin Y^*\times Z^*$, the $2$-path goes to the
    leftover call on $(A,B-B[Y^*,Z^*])$, where $B-B[Y^*,Z^*]$ zeroes out only the entries in $Y^*\times Z^*$.
    If $(y,z)\in Y^*\times Z^*$ and \FindCube{} stopped early because $B[Y^*,Z^*]$ exceeded a constant-factor density
    cap (a \emph{$B$-boosted} return, tagged $\tagdenseB$ in \Cref{sec:overview-depth}), all $2$-paths using $B[Y^*,Z^*]$
    go to a single recursive child on $(A[X,Y^*],B[Y^*,Z^*])$.
    \item \emph{Inside a normal block:} Otherwise, $(x,y)\in X\times Y^*$ belongs to a unique matrix $A_\ell$ in the
    disjoint 2D partition of $A[X,Y^*]$ (either a regular rectangle, an \emph{$A$-boosted} rectangle that exceeded an
    $A$-density cap, or the 2D remainder). For a regular rectangle $A_\ell$, if $z\in Z_\ell$ the $2$-path is output in
    the direct regular piece $(A_\ell,B[Y_\ell,Z_\ell])$; if $z\in R_\ell=Z^*\setminus Z_\ell$, it goes to a recursive
    child on $(A_\ell,B[Y_\ell,R_\ell])$ (which we further partition by column-degree buckets of $Y_\ell$ in
    \Cref{sec:overview-potential}). For the 2D remainder or an $A$-boosted rectangle (where we set $Z_\ell:=\emptyset$ and
    $R_\ell:=Z^*$), all $2$-paths with $z\in Z^*$ go to the corresponding recursive child.
    \item \emph{Base cases:} A sparse call outputs itself as a single sparse piece. At the maximum recursion depth (see
    \Cref{sec:overview-depth}), any remaining nonsparse call partitions the 1-entries of $B$ into disjoint sparse
    \emph{batches} $B_s$ and outputs $(A,B_s)$ for each batch, assigning each $2$-path to the unique batch containing
    $(y,z)$.
\end{enumerate}
Every recursive call either increases the recursion depth or stays at the same depth with strictly fewer $B$-edges, so
the recursion terminates, and induction gives the integer identity $AB=\sum_k A_k B_k$
(\Cref{lem:children-invariants}) without ever multiplying matrices.

\subsection{Where the Improvement Comes From}\label{sec:overview-losses}

A full search has $O(d)$ zooms. If each retained a $2^{-O(d)}$ fraction, the natural bound would already be
$2^{O(d^2)}$. The difficulty is that some zooms are much smaller, pieces overlap, and searches are nested.
Our guiding principle is to charge each loss to its accompanying progress: a density gain, a decrease in weighted
cost, or an amount of depth that is used up. This avoids paying a worst-case loss while crediting only minimum progress.

The four changes in \Cref{tab:bottlenecks}, explained in \Cref{sec:overview-sift,sec:overview-norm,sec:overview-potential,sec:overview-depth} below, show how each of our ideas improves the overall bound.
Idea~2 improves the running time only; the other three account for the piece-count improvement.

\begin{table}[t]
\centering
\footnotesize
\setlength{\tabcolsep}{4pt}
\renewcommand{\arraystretch}{1.08}
\begin{tabularx}{\textwidth}{@{}>{\raggedright\arraybackslash}p{0.17\textwidth}
    >{\raggedright\arraybackslash}X >{\raggedright\arraybackslash}X@{}}
\toprule
\textbf{Bottleneck} & \textbf{Loss in AFKLM~\cite{AFKLM24}} & \textbf{What pays for the loss here} \\
\midrule
1. Sifting
& A single cutoff can retain only $2^{-\Theta(d^2)}$ of the rows at each of $O(d)$ zooms.
& Dyadic levels tie a $2^{-sd}$ loss to a $2^s$ gain; $\sum s\le d$. Removes one power. \\[2pt]
2. Witness search
& Moment accuracy $2^{-\Theta(d^2)}$ leads to long samples and $2^{O(d^3)}$ tuple enumeration.
& Norm estimates, exact row averaging, and a short extractor sampler. Fixes time only. \\[2pt]
3. Query accounting
& $\Theta(d^2)$ raw-volume overhead per level forces pruned $Z$-fraction $\gamma=\Theta(1/d^2)$.
& A quarter-power potential pays for overlap and permits constant $\gamma$. Removes two powers. \\[2pt]
4. Nested recursion
& Full searches inside cube search and across recursion levels compound their losses.
& Constant windows limit ordinary branching; large jumps skip levels. Removes two powers. \\
\bottomrule
\end{tabularx}
\caption{Four changes reduce the piece bound from $2^{O(d^7)}$ to $2^{O(d^2)}$ and make the pieces fast to find.}
\label{tab:bottlenecks}
\end{table}

\subsection{Idea 1: Match Row Loss to Density Gain}\label{sec:overview-sift}

The job of sifting is to find a denser submatrix without making it unnecessarily small.
Suppose the current density is $\mu$ and the matrix has too many $2\times d$ grids. Set
$\delta=(1+\varepsilon/2)\mu$. We first test degrees on the full column set.
If these checks find no rectangle, averaging over the first row $x\in X$ of a $2\times d$ grid yields a \emph{witness
neighborhood} $S=Y_x:=\{y\in Y:A(x,y)=1\}$ with $\abs{S}\ge\delta^2\abs{Y}$ such that the normalized degrees
$Z(x'):=\E_{y\in S}[A(x',y)]$ of rows $x'\in X$ into $S$ have $d$-th moment
$\E_{x'}[Z(x')^d]$ noticeably larger than $\delta^d$ (\Cref{lem:dyadic-sift}).
A large moment means that some rows hit $S$ unusually often. Keeping them with $S$ gives a denser rectangle.

\paragraph{One cutoff does not work.}
The natural choice is to keep every row with $Z\ge\delta$, as in AFKLM.
For illustration only, take $d=10$, far below the algorithm's $d\ge640$, and $\delta$ just above $2^{-10}$.
Suppose a fraction $2\delta^{10}$, on the order of $2^{-100}$, of the rows have $Z=1$, while a $2^{-80}$ fraction
have $Z$ just above $\delta$; the other rows are below the cutoff.
The first group alone supplies more than $\delta^{10}$ to the moment. The second contributes almost nothing
but dominates the selected set. We keep only about $2^{-80}$ of the rows, yet their average degree is still
about $\delta$.

A single cutoff can therefore mix a small spike with many diluting rows, retaining only
$2^{-\Theta(d^2)}$ of the rows for a density gain close to $1+\varepsilon/2$.
Charging this worst-case loss at each of $O(d)$ zooms gives only a $2^{-O(d^3)}$ size guarantee.

\paragraph{Keeping one level.}
We partition the degrees above $\delta$ into power-of-two buckets
\[
    R_s:=\{x':2^s\delta\le Z(x')<2^{s+1}\delta\},
    \qquad 0\le s<d.
\]
They cover all eligible degrees because $2^d\delta>1$. Some bucket carries at least a $1/d$ share of the excess
moment, and each row in it contributes less than $(2^{s+1}\delta)^d$.
Dividing its moment contribution by this per-row bound ties its size to its degree scale.
Concretely, \Sift{} checks buckets from bottom to top and returns the first satisfying
$\abs{R_s}\ge(\varepsilon/16)2^{-(s+2)d}\abs{X}$.
It checks all columns first and, if necessary, the witness neighborhood. If no qualifying bucket exists,
the moment bounds certify regularity (\Cref{alg:sift,lem:dyadic-sift}).

In the toy example, the diluters' bucket is far below its size threshold and is skipped.
The spike's bucket qualifies. Keeping only the spike loses more rows, but raises the density to $1$ and ends the search.

Across a search, let $s_1,\ldots,s_r$ be the bucket scales, assigning scale zero to pruning zooms.
The density can double at most $d$ times, so $\sum_t s_t\le d$, and $r=O(d)$. Apart from fixed constant factors per
step, the retained row fraction is at least
\[
    \prod_{t=1}^r 2^{-(s_t+2)d}
    =2^{-d\sum_t s_t-2rd}
    \ge2^{-O(d^2)}.
\]
Each zoom also retains more than a $2^{-2d}$ fraction of the columns, giving the same cumulative bound there.
These are the size bounds behind \Cref{lem:find-rect,thm:2d-decomp}.
The $d^2$ scale comes from this sifting bound. 

\subsection{Idea 2: Norm Estimation}\label{sec:overview-norm}

The preceding argument must find its witness row deterministically. For each candidate $x$, define its codegrees
and their norm by
\[
    c(x,x'):=\E_{y\in Y}[A(x,y)A(x',y)],
    \qquad
    N(x):=\left(\E_{x'\in X}[c(x,x')^d]\right)^{1/d}.
\]
Since $c(x,x')$ is the fraction of $Y$ adjacent to both $x$ and $x'$, dividing $N(x)$ by the normalized degree
$\abs{Y_x}/\abs{Y}$ of $x$ gives precisely $(\E_{x'}[Z(x')^d])^{1/d}$, the $L_d$ norm of row degrees inside $Y_x$.
We maximize an estimate of this normalized quantity over rows of degree at least $\delta^2$.
Additive error $\alpha'=\varepsilon\delta^3/32$ before division suffices: after division the error is at most
$\varepsilon\delta/32$. At the smallest densities, this requires accuracy $2^{-\Theta(d)}$.
Note that computing every codegree directly amounts to forming $AA^T$, which is the matrix multiplication we are trying to avoid.

\paragraph{Deterministic sampling.}
An $(\varepsilon_{\mathrm{samp}},\delta_{\mathrm{samp}})$-oblivious sampler $\mathcal{T}$ for $Y$ is a deterministic
list of short column sequences $T=(y_1,\dots,y_t)\in Y^t$ of common length $t=\abs{T}\ll\abs{Y}$ such that, for any
function $g:Y\to[0,1]$, the sample mean on $T$ is within $\pm\varepsilon_{\mathrm{samp}}$ of $\E_{y\in Y}[g(y)]$ for at
least a $1-\delta_{\mathrm{samp}}$ fraction of $T\in\mathcal{T}$. Applying this to $g(y)=A(x,y)A(x',y)$ approximates
$c(x,x')$ by $\widehat{c}_T(x,x'):=\frac{1}{t}\sum_{i=1}^t A(x,y_i)A(x',y_i)$, and we estimate $N(x)^d$ by
$\widehat{N}(x)^d:=\E_{T\in\mathcal{T}}\E_{x'\in X}[\widehat{c}_T(x,x')^d]$ (keeping the outer average over $x'\in X$
exact).

Even on a sample $T$, computing $\widehat{c}_T(x,x')$ for every pair $(x,x')\in X^2$ would cost $\Omega(\abs{\mathcal{T}}\abs{X}^2)\gg n^3$.
Instead, we expand $\widehat{c}_T(x,x')^d$ over all $t^d$ index tuples $(i_1,\dots,i_d)\in[t]^d$ and exchange $\E_{x'\in X}$ with the tuple sum:
\[
    \E_{x'\in X}\bigl[\widehat{c}_T(x,x')^d\bigr]
    = \frac{1}{t^d}\sum_{(i_1,\dots,i_d)\in[t]^d}
      \Bigl(\prod_{j=1}^d A(x,y_{i_j})\Bigr)
      \Bigl(\E_{x'\in X}\prod_{j=1}^d A(x',y_{i_j})\Bigr).
\]
For each $T\in\mathcal{T}$ and tuple $(i_1,\dots,i_d)\in[t]^d$, one scan of $X$ computes the fraction of rows $x'\in X$
adjacent to all $d$ columns $y_{i_1},\dots,y_{i_d}$, and a second scan of $X$ adds that single number to every candidate
row $x\in X$ also adjacent to all $d$ columns. This drops $\abs{X}^2$ to $\abs{X}$, for total work proportional to
$\abs{\mathcal{T}}\abs{T}^d\abs{X}$ (up to factors for $d$ and exact arithmetic). Notice that $\abs{\mathcal{T}}$ enters
linearly, whereas sequence length $\abs{T}$ is raised to the $d$-th power: $\abs{T}=\poly(\log n)\,2^{O(d)}$ gives
$\abs{T}^d=(\log n)^{O(d)}2^{O(d^2)}$, whereas $\abs{T}=2^{\Theta(d^2)}$ blows up to $\abs{T}^d=2^{\Theta(d^3)}$.

\paragraph{Accuracy and failures.}
Fine accuracy can force length on the order of $1/\varepsilon_{\mathrm{samp}}$, whereas suitable samplers need
length only polynomial in $\log(1/\delta_{\mathrm{samp}})$ for high confidence.
Accuracy $2^{-\Theta(d)}$ suffices for our norm gap. Failures, on the other hand, are more dangerous: a failed sequence might report
a codegree near $1$ instead of $2^{-2d}$, contributing $1$ instead of $2^{-2d^2}$ after the $d$-th power.
Even a small fraction of failures can overwhelm the true moment.

For fixed $x$, let $a(x',T)$ be the sampled codegree and $b(x',T)=c(x,x')$ the true one.
Their $L_d$ norms on $X\times\mathcal{T}$ are the estimate $\widehat N(x)$ and the true $N(x)$.
The triangle inequality gives
\[
    \abs{\widehat N(x)-N(x)}
    \le \norm{a-b}_{L_d(X\times\mathcal{T})}
    \le \bigl(\varepsilon_{\mathrm{samp}}^d+\delta_{\mathrm{samp}}\bigr)^{1/d}
    \le \varepsilon_{\mathrm{samp}}+\delta_{\mathrm{samp}}^{1/d}.
\]
We may take $\varepsilon_{\mathrm{samp}}=\alpha'/2=2^{-O(d)}$ and
$\delta_{\mathrm{samp}}=(\alpha'/2)^d=2^{-O(d^2)}$.
There is no union bound over $x'$: different rows may fail on different sequences, but bad pairs $(x',T)$ occupy
at most a $\delta_{\mathrm{samp}}$ fraction of the product space. No single sequence must work for every row.

AFKLM estimated the unrooted $d$-th-power quantity, whose relevant gaps can be $2^{-\Theta(d^2)}$.
This placed the tiny parameter in the accuracy requirement, making samples too long.
The exact outer averaging and the norm inequality we use avoid this problem, but a second change is needed:
their Goldreich--Wigderson sampler~\cite{GoldreichW97} has length polynomial in
$1/\delta_{\mathrm{samp}}$, which would still be too long.

We instead use Zuckerman's sampler--extractor correspondence with the GUV extractor~\cite{Zuckerman97,GUV09,Vadhan12}.
The sampler of \Cref{prop:sampler} has
$\abs{\mathcal{T}}=O(\abs{Y}^{5/4}\varepsilon_{\mathrm{samp}}^{-5/4}\delta_{\mathrm{samp}}^{-1})$ and
$\abs{T}\le\poly(\log(\abs{Y}+2),\varepsilon_{\mathrm{samp}}^{-1},\log(2/\delta_{\mathrm{samp}}))$.
At these parameters, the family size is $\abs{Y}^{5/4}2^{O(d^2)}$, but sequence length is only
$\poly(\log n)\,2^{O(d)}$; only the latter is raised to the $d$-th power.
Accounting for the $(\log n)^{O(d)}$ factor and exact arithmetic gives
$O(n^{5/2}2^{O(d^2)})$ time for \EstimateNorm{} (\Cref{lem:approx-grid-norm}).
Note that the $d$-th roots are only for analysis: the algorithm compares exact rational $d$-th powers.

\subsection{Idea 3: Weight Pieces by Their Query Cost}\label{sec:overview-potential}

We now bound the potential that pays for the queries.
Note that the overlap in the pieces already appears in peeling: consider $d$ edge-disjoint pieces, each covering half the original area,
with densities $1/2,1/4,\ldots,2^{-d}$. Their ones fit easily, but their areas sum to $d/2$ times the original area.
As the leftover becomes sparser, a region can be covered again at a lower density.
In three dimensions this occurs on both sides and can give $\Theta(d^2)$ overcounting per level, compounded through
calls on omitted vertices. AFKLM therefore used $\gamma=\Theta(1/d^2)$ to make those children small enough,
costing two powers of $d$ in the search.

\paragraph{Using weighted area.}
Weighting by query cost makes the overlap summable.
In the example, weighting each rectangle by its density to the $1/4$ power replaces $d/2$ by
$\frac12\sum_i2^{-i/4}=O(1)$.

For each extraction, let $t_\ell$ be the leftover density before rectangle $\ell$, let $\mu_\ell\ge t_\ell$
be that rectangle's density, and let $a_\ell$ be its fraction of the original area.
Peeling lowers the density by $t_\ell-t_{\ell+1}=a_\ell\mu_\ell$. Hence
\[
    a_\ell\le\frac{t_\ell-t_{\ell+1}}{t_\ell},
    \qquad
    a_\ell\mu_\ell^{1/4}
       \le\frac{t_\ell-t_{\ell+1}}{t_\ell^{3/4}}.
\]
The first sum costs $O(d)$ via $\int_{2^{-d}}^1dt/t$, plus at most $1$ for the last extraction.
The second is bounded by $\int_0^{t_1}t^{-3/4}\,dt$, which converges.
Including the final leftover, the weighted areas total at most
$4\dens{A}^{1/4}\abs{X}\abs{Y}$ (\Cref{lem:riemann}).

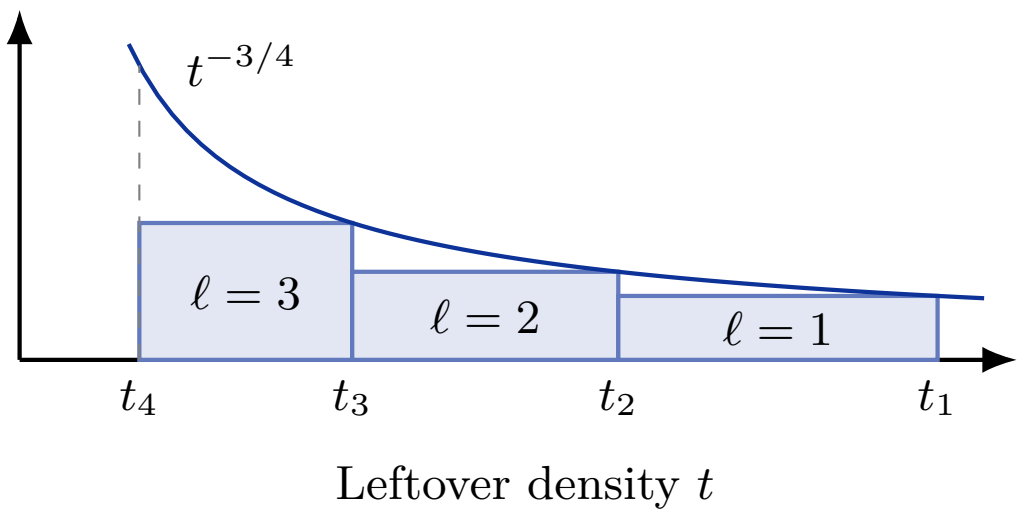
\begin{figure}[t]
\centering
\begin{tikzpicture}[x=0.88cm,y=0.8cm,>=Latex,font=\small]
    \draw[->,thick] (0,0) -- (7.5,0);
    \draw[->,thick] (0,0) -- (0,2.9);
    \fill[darkblue!12,draw=darkblue!65,thick] (0.9,0) rectangle (2.5,{2.25/(2.5^0.75)});
    \fill[darkblue!12,draw=darkblue!65,thick] (2.5,0) rectangle (4.5,{2.25/(4.5^0.75)});
    \fill[darkblue!12,draw=darkblue!65,thick] (4.5,0) rectangle (6.9,{2.25/(6.9^0.75)});
    \draw[darkblue,thick,domain=0.82:7.25,samples=60] plot (\x,{2.25/(\x^0.75)});
    \draw[dashed,gray] (0.9,0) -- (0.9,{2.25/(0.9^0.75)});
    \node[anchor=south west,font=\footnotesize] at (1.1,{2.25/(1.1^0.75)}) {$t^{-3/4}$};
    \node[below,font=\footnotesize] at (0.9,0) {$t_4$};
    \node[below,font=\footnotesize] at (2.5,0) {$t_3$};
    \node[below,font=\footnotesize] at (4.5,0) {$t_2$};
    \node[below,font=\footnotesize] at (6.9,0) {$t_1$};
    \node[font=\footnotesize] at (1.7,0.56) {$\ell=3$};
    \node[font=\footnotesize] at (3.5,0.36) {$\ell=2$};
    \node[font=\footnotesize] at (5.7,0.26) {$\ell=1$};
    \node[font=\footnotesize] at (3.8,-1.05) {Leftover density $t$};
\end{tikzpicture}
\caption{Each weighted area fraction is bounded by a box below $t^{-3/4}$. The boxes occupy disjoint density
intervals, so their sum is at most $4t_1^{1/4}$.}
\label{fig:overview-riemann}
\end{figure}

\paragraph{A carefully chosen budget.}
A recursive call has depth $h$ and a parameter $\alpha\in(0,1]$ such that every column $y\in Y$ has normalized degree
$\E_{x\in X}[A(x,y)]\le 2\alpha$ in $A$. With $\beta:=\dens{B}$, we give its output the budget
\[
    \Lambda_h\alpha^{1/4}\beta^{1/4}\abs{X}\abs{Y}\abs{Z},
    \qquad \Lambda_h:=1024\cdot2^h.
\]
We use this cap, not $\dens{A}$, because it survives restriction to \emph{any} column subset $Y^*$:
$\dens{A[X,Y^*]}\le2\alpha$. An average-density bound would not prevent the search from selecting unusually heavy
columns.

Row restrictions can break the cap, so before recursing on $A_\ell$ we bucket its columns at caps
$2^{j+1}\mu$, where $\mu:=\dens{A_\ell}>2^{-d}$.
Markov's inequality bounds bucket $j$'s fraction by $2^{-j}$; together with conservation of $B$-edges, this limits
the weighted overhead to $\sum_{j\ge0}2^{-j/2}<7/2$ (\Cref{lem:dyadic-bucketing}).

A sparse call outputs one piece of weight at most
$(2\alpha)^{1/4}\beta^{1/4}\abs{X}\abs{Y}\abs{Z}$, well below its budget.
At the depth ceiling $H=\lceil3d/4\rceil$, a nonsparse call partitions $B$'s ones on $Y\times Z$ into $N<3\cdot2^d$
disjoint batches $B_s$ of density at most $2^{-d}$; since $\sum_s\dens{B_s}=\beta$, H\"older's inequality gives
$\sum_s\dens{B_s}^{1/4}\le N^{3/4}\beta^{1/4}$.
The extra factor is only a constant times $2^{3d/4}\le2^H$, which is already allowed by $\Lambda_H$.

\paragraph{Deleting blocks and refunds.}
We prove by induction that every call $(X,Y,Z,A,B,\alpha,h)$ outputs pieces $\mathcal{S}$ with
$\pot(\mathcal{S})\le\mathcal{B}:=\Lambda_h\alpha^{1/4}\beta^{1/4}\abs{X}\abs{Y}\abs{Z}$ (where $\beta:=\dens{B}$); at
the root ($h=0,\alpha=1,\Lambda_0=1024$), this gives $\pot(\mathcal{S})\le 1024\,n^3$.
The key question is how a parent call can pay for \emph{both} its leftover call on $(X,Y,Z,A,B-B[Y^*,Z^*],\alpha,h)$---which
still lives on the full vertex sets $X,Y,Z$ at depth $h$---and all pieces and recursive calls inside the extracted block
$Y^*\times Z^*$. The unweighted volume does not work here, since deleting edges does not shrink $X,Y,Z$.

With $\beta^{1/4}$ weighting, deleting $B[Y^*,Z^*]$ lowers $B$'s density and frees up budget. Let
$a:=\abs{Y^*}\abs{Z^*}/(\abs{Y}\abs{Z})$ be the block's area fraction and $\beta^*:=\dens{B[Y^*,Z^*]}\ge\beta$ its
density. Removing the block deletes a fraction $u:=a\beta^*/\beta$ of $B$'s ones, leaving leftover density
$\beta_{\mathrm{res}}=\beta(1-u)$. By concavity, $(1-u)^{1/4}\le 1-u/4$, so
$\beta_{\mathrm{res}}^{1/4}\le\beta^{1/4}-\frac14(\beta^*/\beta)^{3/4}a(\beta^*)^{1/4}$. Multiplying by
$\Lambda_h\alpha^{1/4}\abs{X}\abs{Y}\abs{Z}$ and writing
$W^*:=\Lambda_h\alpha^{1/4}(\beta^*)^{1/4}\abs{X}\abs{Y^*}\abs{Z^*}$ for the block's budget scale, the leftover call's
inductive budget drops below $\mathcal{B}$ by a \emph{refund} of at least
\[
    \frac14\left(\frac{\beta^*}{\beta}\right)^{3/4}W^*
    \ge \frac14 W^*.
\]

To see why this $\frac14 W^*$ refund covers everything inside $Y^*\times Z^*$, first ignore density caps and suppose
\FindCube{} decomposes $A[X,Y^*]$ into regular rectangles $A_\ell$ (plus a sparse remainder):
\begin{itemize}[leftmargin=*,nosep,topsep=2pt,itemsep=2pt]
    \item \emph{Direct pieces $(A_\ell, B[Y_\ell,Z_\ell])$:} These are output immediately rather than recursed on, so
    their weight has \emph{no} $\Lambda_h$ multiplier. Using the 2D weighted-area bound
    $\sum_\ell\dens{A_\ell}^{1/4}\abs{X_\ell}\abs{Y_\ell}\le 4\dens{A[X,Y^*]}^{1/4}\abs{X}\abs{Y^*}$
    (\Cref{lem:riemann}) and $\dens{B[Y_\ell,Z_\ell]}\le 3\beta^*$, their total weight is only $O(1/\Lambda_h)\,W^*$,
    which is tiny once $\Lambda_0=1024$.
    \item \emph{Omitted $Z$-vertices ($R_\ell=Z^*\setminus Z_\ell$):} Recursing one level deeper doubles $\Lambda_h$,
    and column bucketing loses only a constant factor ($\sum_{j\ge 0}2^{-j/2}<7/2$, \Cref{lem:dyadic-bucketing}), multiplying the budget scale by $O(1)$. Because pruning omitted fewer than
    $\gamma\abs{Z^*}$ vertices ($\abs{R_\ell}<\gamma\abs{Z^*}$) and the same 2D weighted-area bound, when summed over $\ell$, 
    is within a factor of $4$, their total budget is $O(\gamma)\,W^*$, which is tiny for a small constant $\gamma=1/4096$.
\end{itemize}

In \Cref{sec:overview-depth} (Idea~4), we also stop a search early whenever the density leaves a large constant-factor
window ($[2^{-40}\alpha,2^{40}\alpha]$ for $A$, or above $2^{40}\beta$ for $B$) and recurse on that piece. Both early
exits easily fit inside the refund (\Cref{tab:potential-budget,lem:potential}):
\begin{itemize}[leftmargin=*,nosep,topsep=2pt,itemsep=2pt]
    \item \emph{Below the floor ($\dens{A_L}\le 2^{-40}\alpha$):} Recursing on the nonsparse 2D remainder $A_L$ pays the
    constant recursion and bucketing overhead, which is overwhelmed by $(2^{-40})^{1/4}=2^{-10}$.
    \item \emph{Above the cap (density jump $\rho\ge 2^{40}$):} When $B$'s density jumps by $\rho=\beta^*/\beta\ge 2^{40}$,
    the refund above grows to $\frac14\rho^{3/4}W^*$, whereas Idea~4 advances the depth by $\Delta h\le\frac12\log_2\rho$,
    multiplying $\Lambda_h$ by at most $\rho^{1/2}\ll\frac14\rho^{3/4}$ (since $\rho^{1/4}\ge 2^{10}$). The same gap
    $\rho_\ell^{1/2}\ll\rho_\ell^{3/4}$, summed across edge-disjoint rectangles $A_\ell$, pays for high-density
    $A$-rectangles ($\rho_\ell=\dens{A_\ell}/\alpha\ge 2^{40}$).
\end{itemize}

Because the omitted-$Z$ overhead before $\gamma$ is $O(1)$ rather than AFKLM's $\Theta(d^2)$, we can take constant $\gamma=1/4096$ instead of $\Theta(1/d^2)$, so each failed prune raises $B$'s density by $\ge 1+\varepsilon\gamma/4$ and a 
constant-factor window allows only $O(1)$ restarts. Finally, the exponent $p=1/4$ sits in $(0,1/2)$: $p>0$ bounds the 2D weighted-area sum by
$\frac{1}{p}\dens{A}^p\abs{X}\abs{Y}$ (\Cref{lem:riemann}), $p<1$ gives concavity for the refund, and $p<1/2$ gives
both the Case~2 query saving and a convergent column-bucketing sum $\sum_{j\ge 0}2^{j(2p-1)}$.

\subsection{Idea 4: Use Depth to Pay for Branching}\label{sec:overview-depth}

The remaining job is to bound the entire recursion tree.
There are two dangers. Children on omitted vertices shrink a side only by a constant factor, so without some ceiling
they can nest for $\Theta(\log n)$ levels; a $2^{O(d)}$ branching per level would then give a prohibitive $n^{O(d)}$ pieces.
But also limiting the depth to $O(d)$ is not enough: a full $2^{O(d^2)}$ search at every level could still give
$2^{O(d^3)}$ pieces.

\paragraph{Capping ordinary branching at $2^{O(d)}$ per level, and skipping levels on large jumps.}
A natural question is why an unrestricted search can produce $2^{O(d^2)}$ rectangles. This happens only when its zooms
multiply density by a large factor $2^s$ (up to $2^d$), losing a $2^{-O(sd)}$ fraction of the vertices along the way.
We therefore stop an ordinary search as soon as its density changes by more than a large constant factor $C=O(1)$
(we take $C=2^{40}$: outside $[C^{-1}\alpha,C\alpha]$ for $A$, or above $C\beta$ for $B$).
Inside this constant-factor window, the density can grow by only $O(1)$, so a search makes only $O(1)$ zooms, every regular
piece retains at least a $2^{-O(d)}$ fraction of the area, and ordinary peeling produces only $2^{O(d)}$ direct pieces
and $2^{O(d)}$ depth-$(h+1)$ children.

What if a search \emph{does} jump above its cap, attaining a large density ratio $\rho=\dens{A_\ell}/\alpha\ge C$
($\tagdenseA$) or $\rho=\beta^*/\beta\ge C$ ($\tagdenseB$)? Rather than keep zooming, we stop immediately and recurse on
that block at depth $h+\Delta h$, advancing the explicit depth $h\le H:=\lceil 3d/4\rceil$ by
\[
    \Delta h:=\min\!\left(H-h,\left\lfloor\tfrac12\log_2\rho\right\rfloor\right).
\]
Intuitively, a density gain of roughly $2^s$ can shrink the block's area to $2^{-O(sd)}$---allowing up to $2^{O(sd)}$
such blocks---but it also consumes $\Delta h=\Theta(s)$ levels of depth at once (unless it hits the ceiling $H$). Thus
the branching paid \emph{per level of depth consumed} is still $(2^{O(sd)})^{1/\Theta(s)}=2^{O(d)}$.

\paragraph{Why batching works at the ceiling $H=\lceil 3d/4\rceil$, though not at the root.}
When a call reaches the depth ceiling $h=H$ without becoming sparse, we simply partition $B$'s ones into $N<3\cdot 2^d$
disjoint batches of density at most $2^{-d}$ (as in Idea~3). Doing this at the root ($h=0$) would be useless---it merely
splits a brute-force scan of $B$ into pieces and, by H\"older's inequality, inflates the weighted volume by
$N^{3/4}=O(2^{3d/4})$, wiping out our query saving. By depth $H=\lceil 3d/4\rceil$, however, the budget multiplier
$\Lambda_H=1024\cdot 2^H\ge 1024\cdot 2^{3d/4}$ has already been paid for by the density drops, small $\gamma$-fractions,
and $\rho^{3/4}$-vs.-$\rho^{1/2}$ margins along the path from the root (\Cref{sec:overview-potential}), so the
accumulated subproblem is already small enough to absorb the $2^{3d/4}$ batching factor.

\paragraph{Solving the tree recurrence.}
Consider the sequence of leftover calls $(A,B-B[Y^*,Z^*])$ at a fixed depth $h$ before $B$ becomes sparse. Let
$S_B\in\{0,\dots,d\}$ be the sum of sifting scales inside \FindCube{}. At $S_B=s$, the extracted block retains at least a
$2^{-O(d)-sd}$ fraction of $Y\times Z$ at density $>2^{-d}$ (\Cref{lem:cube-step}), so zeroing it out removes at least a
$2^{-O(d)-sd}$ fraction of $\abs{Y}\abs{Z}$ entries; hence at most $2^{O(d)+sd}$ extractions along the depth-$h$ chain
can have $S_B=s$.
Normal extractions ($\tagnormal$) stay inside the constant-factor window, so $S_B=O(1)$, giving $2^{O(d)}$ direct pieces
and $2^{O(d)}$ depth-$(h+1)$ children across the entire chain. Meanwhile, at sifting scale $s$, there are at most
$2^{O(d)+sd}$ boosted children ($\tagdenseA$ or $\tagdenseB$) along the chain, and each advances in depth by at least
$D(s):=\min(H-h,\Theta(s))$ (with $D(s)\ge 1$; see \Cref{lem:recurrence} for exact constants).

Let $T(h)$ be the maximum number of pieces output by any call at depth at least $h$. Summing over the chain gives
\[
    T(h)\le 2^{O(d)}\bigl(1+T(h+1)\bigr)
       +\sum_{s=0}^d 2^{O(d)+sd}\,T\bigl(h+D(s)\bigr),
    \qquad T(H)\le3\cdot2^d.
\]
By downward induction, $T(h)\le 2^{d^2+O(d)(H-h)}$:
\begin{itemize}[leftmargin=*,nosep,topsep=2pt,itemsep=2pt]
    \item If a jump stops \emph{below} the ceiling ($h+D(s)<H$), then $D(s)=\Theta(s)$, so $2^{sd}=(2^{O(d)})^{D(s)}$
    costs only $2^{O(d)}$ per skipped level, matching the $2^{O(d)(H-h)}$ rate.
    \item If a jump lands \emph{at} the ceiling ($h+D(s)=H$), then $T(H)\le 3\cdot 2^d$ has no $2^{d^2}$ factor yet, so
    $2^{sd}T(H)\le 3\cdot 2^{d^2+d}$ pays the $2^{d^2}$ factor at most once, at the leaf.
\end{itemize}
Thus at the root we obtain $K\le T(0)=2^{O(d^2)}$ pieces (\Cref{lem:recurrence}). Since each call to \FindCube{} takes
$O(n^{5/2}2^{O(d^2)})$ time and there are $2^{O(d^2)}$ calls across the tree, total preprocessing time is
$O(n^{5/2}2^{O(d^2)})$.

\subsection{Putting It Together}\label{sec:overview-finish}

Choose $d\ge640$ as a multiple of $320$ (so $q=2^{-d/320}$ is an integer power of $1/2$) with $d=\Theta(\sqrt{\log n})$ and
a sufficiently small leading constant so that $K\le 2^{O(d^2)}=O(n^{1/4})$. Preprocessing then takes
$O(n^{5/2}K)=O(n^{11/4})$ time, and the query phase takes
$O(n^2K+n^3/2^{d/640})=O(n^3/2^{\Omega(\sqrt{\log n})})$ time.
For triangle detection in an arbitrary graph $G=(V,E)$, the standard three-copy construction creates three copies
$X,Y,Z$ of $V$ and places the adjacency matrix of $G$ between each pair of copies.
Finally, the combinatorial reduction of \Cref{prop:bmm-to-tridet} reduces $n\times n$ BMM to tripartite triangle
detection on graphs of size $N\ge\lfloor n^{1/3}\rfloor$, turning an $O(N^3/f(N))$ triangle algorithm into an
$O(n^3/f(\lfloor n^{1/3}\rfloor))$ BMM algorithm; since $\sqrt{\log(n^{1/3})}=\Theta(\sqrt{\log n})$, this changes only
the hidden constant in the  $\Omega(\sqrt{\log n})$.

\paragraph{Proof roadmap.}
\Cref{lem:approx-grid-norm,lem:dyadic-sift} give our norm estimation and sifting;
\Cref{lem:find-rect,thm:2d-decomp,lem:riemann} give rectangle search, peeling, and weighted-area bounds.
\Cref{lem:dyadic-bucketing,lem:cube-step} give the column caps and aligned cube search, and
\Cref{lem:children-invariants,lem:potential,lem:recurrence} prove exactness, the potential bound, and the tree count.
The three-case query is in \Cref{sec:bmm}.
The black boxes used above are KLM, AFKLM's pruning procedure, the sampler, and the BMM reduction
(\Cref{prop:regular-product,lem:min-degree,prop:sampler,prop:bmm-to-tridet}).
Sampling details, deferred proofs, and exact integer implementation appear in
\Cref{app:sampling,app:helper-proofs,app:applications}.

\clearpage
\section{Preliminaries}\label{sec:prelim}

\subsection{Notation and Combinatorial Model}\label{sec:notation}

We work in the standard $O(\log n)$-bit Word RAM model restricted to elementary integer arithmetic ($+, -, \times, \div$), comparisons, and array indexing---using no bitwise operations, bit packing, floating-point arithmetic, or Four-Russians lookup tables~\cite{AFKLM24}. Exact rationals in $\EstimateNorm$ (\Cref{sec:2d-norm}) have $O(d\log n + d^2)$-bit integer numerators and denominators and are manipulated via standard multiprecision integer arithmetic, included in all running times (\Cref{app:sampling}).

We set $[0] := \emptyset$ and $[k] := \{1, \dots, k\}$ for integers $k \ge 1$, and write $\E_{x \in S}$ and $\Pr_{x \in S}$ for expectation and probability under the uniform distribution on a finite non-empty set $S$. Unless stated otherwise, all vertex sets $X, Y, Z$ are finite and non-empty, and we fix $\varepsilon := 1/160$ and $\gamma := 2^{-12} = 1/4096$.
\begin{itemize}[nosep, leftmargin=*, topsep=2pt]
    \item \textbf{Support, density, and submatrices.}
    For $A \in \R_{\ge 0}^{X \times Y}$, define $\supp(A) := \{(x, y) \in X \times Y : A(x, y) \ne 0\}$, $\nnz(A) := \abs{\supp(A)}$, and $\dens{A} := \E_{(x, y) \in X \times Y}[A(x, y)]$. For non-empty $X' \subseteq X$ and $Y' \subseteq Y$, $A[X', Y']$ denotes the induced submatrix (with expectations uniform over $X' \times Y'$); in ambient sums or differences ($A - A[X', Y']$, $\sum_\ell A_\ell$), submatrices are zero-extended outside their index sets.

    \item \textbf{Degrees, neighborhoods, and $\varepsilon$-min-degree.}
    Set $\deg_A(x) := \E_{y \in Y}[A(x, y)]$ and $\deg_{A^T}(y) := \E_{x \in X}[A(x, y)]$. For binary $A \in \{0, 1\}^{X \times Y}$ and $x \in X$, let $Y_x := \{y \in Y : A(x, y) = 1\}$ and $A_x := A[X, Y_x]$ when $\deg_A(x) > 0$, with $\deg_A(x)^d\norm{A_x}_{U(1, d)}^d := 0$ when $\deg_A(x) = 0$. A matrix $A \in \R_{\ge 0}^{X \times Y}$ is \emph{$\varepsilon$-min-degree} if $\deg_A(x) \ge (1 - \varepsilon)\dens{A}$ for all $x \in X$.

    \item \textbf{Matrix products and uniformity.}
    For $A \in \R_{\ge 0}^{X \times Y}$ and $B \in \R_{\ge 0}^{Y \times Z}$, set $(AB)(x, z) := \sum_{y \in Y} A(x, y)B(y, z)$ and $(A \circ B)(x, z) := \E_{y \in Y}[A(x, y)B(y, z)] = \frac{1}{\abs{Y}}(AB)(x, z)$; for binary $A, B$, $(A \star B)(x, z) := \bigvee_{y \in Y}(A(x, y) \wedge B(y, z))$. A matrix $M \in \R_{\ge 0}^{X \times Z}$ is \emph{$(\alpha, \varepsilon, \delta)$-uniform} if $\Pr_{(x, z) \in X \times Z}[(1 - \varepsilon)\alpha \le M(x, z) \le (1 + \varepsilon)\alpha] \ge 1 - \delta$.
\end{itemize}

\subsection{Grid Norms and Regularity}\label{sec:grid-norms}

\begin{definition}[Grid norm and regularity~{\cite{KelleyLM23,AFKLM24}}]\label{def:grid-norm}
For $A \in \R_{\ge 0}^{X \times Y}$ on finite non-empty sets $X, Y$ and integers $k, \ell \ge 1$, let $\mathbf{x} = (x_1, \dots, x_k) \in X^k$ and $\mathbf{y} = (y_1, \dots, y_\ell) \in Y^\ell$ be independent uniform tuples, and define $\norm{A}_{U(k, \ell)} \ge 0$ via Fubini's theorem by
\begin{equation}\label{eq:grid-norm-fubini}
    \norm{A}_{U(k, \ell)}^{k\ell}
    := \E_{\mathbf{x},\, \mathbf{y}} \prod_{i=1}^k \prod_{j=1}^\ell A(x_i, y_j)
    = \E_{\mathbf{x}} \Bigl( \E_{y \in Y} \prod_{i=1}^k A(x_i, y) \Bigr)^\ell
    = \E_{\mathbf{y}} \Bigl( \E_{x \in X} \prod_{j=1}^\ell A(x, y_j) \Bigr)^k.
\end{equation}
For $\varepsilon > 0$, we say $A$ is \emph{$(\varepsilon, k, \ell)$-regular} if $\norm{A}_{U(k, \ell)} \le (1 + \varepsilon)\dens{A}$.
\end{definition}

\begin{proposition}[Uniformity of regular products~{\cite[Lemma~4.8]{KelleyLM23}; see also~\cite[Theorem~2.1]{AFKLM24}}]\label{prop:regular-product}
Let $A \in \R_{\ge 0}^{X \times Y}$ and $B \in \R_{\ge 0}^{Y \times Z}$ be non-zero matrices on finite non-empty sets $X, Y, Z$, let $\varepsilon \in (0, 1/80)$, and let $d \ge 4/\varepsilon$ be an even integer. If $A$ and $B^T \in \R_{\ge 0}^{Z \times Y}$ are both $(\varepsilon, 2, d)$-regular and $\varepsilon$-min-degree, then $A \circ B$ is $(\dens{A}\dens{B}, \, 80\varepsilon, \, 2^{-\varepsilon d / 2})$-uniform on $X \times Z$. In particular, for $\varepsilon = 1/160$ and even $d \ge 640$, at most a $2^{-\varepsilon d / 2}$-fraction of pairs $(x, z) \in X \times Z$ satisfy $(AB)(x, z) = 0$.
\end{proposition}

\subsection{Minimum-Degree Pruning and Oblivious Sampling}\label{sec:prelim-tools}

\begin{proposition}[Minimum-degree pruning~{\cite[Lemma~5.1]{AFKLM24}}]\label{lem:min-degree}
For finite non-empty sets $U, V$, binary $M \in \{0, 1\}^{U \times V}$, $\varepsilon_0 \in (0, 1)$, and $\gamma_0 \in (0, 1/2]$, a deterministic $O(\abs{U}\abs{V})$-time greedy row-deletion procedure $\Prune(U, V, M, \varepsilon_0, \gamma_0)$ returns a non-empty subset $U' \subseteq U$ with $\abs{U'} \ge \max(1, \lfloor(1 - \gamma_0)\abs{U}\rfloor) \ge \frac{1}{3}\abs{U}$ and a status flag (\textnormal{\emph{certified}} or \textnormal{\emph{boosted}}) such that, for $M' := M[U', V]$:
\begin{enumerate}[nosep, leftmargin=*, topsep=2pt]
    \item \textbf{Certified:} $M'$ is $\varepsilon_0$-min-degree on $U'$, $\dens{M'} \ge \dens{M}$, $\abs{U \setminus U'} < \gamma_0\abs{U}$, and if $U \setminus U' \ne \emptyset$ then $\dens{M[U \setminus U', V]} < (1 - \varepsilon_0)\dens{M'}$.
    \item \textbf{Boosted:} $\abs{U} \ge 2$, $\abs{U'} = \lfloor(1 - \gamma_0)\abs{U}\rfloor$ (so $\abs{U \setminus U'} \ge \gamma_0\abs{U}$), and $\dens{M'} \ge \frac{1}{1 - \gamma_0\varepsilon_0}\dens{M} \ge (1 + \gamma_0\varepsilon_0)\dens{M}$.
\end{enumerate}
\end{proposition}

For finite non-empty $Y$ and $\varepsilon_{\mathrm{samp}}, \delta_{\mathrm{samp}} \in (0, 1)$, an \emph{$(\varepsilon_{\mathrm{samp}}, \delta_{\mathrm{samp}})$-oblivious sampler} $\mathcal{T}$ of $Y$ is a non-empty finite indexed family of non-empty sequences $T \in Y^{\abs{T}}$ such that for every $g : Y \to [0, 1]$,
\begin{equation}\label{eq:oblivious-sampler}
    \Pr_{T \in \mathcal{T}}\!\left[\,\abs{\E_{y \in T}[g(y)] - \E_{y \in Y}[g(y)]} \le \varepsilon_{\mathrm{samp}}\,\right] \ge 1 - \delta_{\mathrm{samp}},
\end{equation}
where $\E_{y \in T}[g(y)] := \frac{1}{\abs{T}}\sum_{i=1}^{\abs{T}} g(T(i))$ and $T^d$ is uniform on~$[\abs{T}]^d$. Combining Zuckerman's sampler reduction~\cite{Zuckerman97} with the Guruswami--Umans--Vadhan extractor~\cite{GUV09} (\cite[Corollary~6.24 and Theorem~6.36]{Vadhan12}) and power-of-two domain padding yields short sample sequences (proved in \Cref{app:sampling}, along with \Cref{rem:sampler-choice}):

\begin{proposition}[Extractor-based oblivious sampler]\label{prop:sampler}
For every finite non-empty set $Y$ and $\varepsilon_{\mathrm{samp}}, \delta_{\mathrm{samp}} \in (0, 1)$, one can deterministically construct in time $\abs{Y}^{5/4}\poly(\log(\abs{Y} + 2), \varepsilon_{\mathrm{samp}}^{-1}, \delta_{\mathrm{samp}}^{-1})$ an $(\varepsilon_{\mathrm{samp}}, \delta_{\mathrm{samp}})$-oblivious sampler $\mathcal{T}$ of $Y$ of equal-length sequences $T$ satisfying
\[
    \abs{\mathcal{T}} = O\!\left(\abs{Y}^{5/4} \varepsilon_{\mathrm{samp}}^{-5/4} \delta_{\mathrm{samp}}^{-1}\right), \qquad
    \abs{T} \le \poly\!\left(\log(\abs{Y} + 2),\, \varepsilon_{\mathrm{samp}}^{-1},\, \log(2/\delta_{\mathrm{samp}})\right).
\]
\end{proposition}

\section{Two-Dimensional Regularity Decomposition}\label{sec:2d}

Throughout this section, $A \in \{0, 1\}^{X \times Y}$ is a binary matrix on finite non-empty sets $X, Y$ ($n := \abs{X} + \abs{Y} \ge 2$), with $\varepsilon = 1/160$, integer $d \ge 1$ in \Cref{sec:2d-norm}, and integer $d \ge 2$ in \Cref{sec:2d-sift,sec:2d-decomp}.

\subsection{Deterministic Grid-Norm Estimation}\label{sec:2d-norm}

Setting $A'_x(x', y) := A(x', y)A(x, y)$ gives $\norm{A'_x}_{U(1, d)} = \deg_A(x)\norm{A_x}_{U(1, d)}$. Our procedure $\EstimateNorm(X, Y,\allowbreak A,\allowbreak d, \alpha')$ averages over $X$ \emph{exactly} once per sampled tuple and returns exact rationals $V_x = v_x^d$:
\begin{enumerate}[nosep, leftmargin=*, topsep=2pt]
    \item Construct an $(\varepsilon_{\mathrm{samp}}, \delta_{\mathrm{samp}})$-oblivious sampler $\mathcal{T}$ of $Y$ (\Cref{prop:sampler}) with $\varepsilon_{\mathrm{samp}} := \alpha'/2$ and $\delta_{\mathrm{samp}} := (\alpha'/2)^d$.
    \item For each $T \in \mathcal{T}$ and $\mathbf{y} = (y_1, \dots, y_d) \in T^d$, set $u_{T, \mathbf{y}} := \E_{x' \in X}\bigl[\prod_{j=1}^d A(x', y_j)\bigr]$.
    \item For each $x \in X$ with $\deg_A(x) > 0$, set $U_x := \E_{T \in \mathcal{T}}\E_{\mathbf{y} \in T^d}\bigl[u_{T, \mathbf{y}}\prod_{j=1}^d A(x, y_j)\bigr]$ and return $V_x := U_x / \deg_A(x)^d \ge 0$.
\end{enumerate}

\begin{lemma}[Grid-norm estimation]\label{lem:approx-grid-norm}
For integer $d \ge 1$ and $\alpha' \in (0, 1)$, $\EstimateNorm(X, Y,\allowbreak A,\allowbreak d, \alpha')$ runs in deterministic time $O(n^{5/2}(d/\alpha')^{O(d)})$ and outputs rationals $V_x \ge 0$ for all $x \in X$ with $\deg_A(x) > 0$ such that $v_x := V_x^{1/d}$ obeys
\begin{equation}\label{eq:approx-grid-norm-err}
    \abs{v_x - \norm{A_x}_{U(1, d)}} \le \frac{\alpha'}{\deg_A(x)}.
\end{equation}
\end{lemma}

\begin{proof}
Fix $x \in X$ with $\deg_A(x) > 0$. Writing $\norm{\cdot}_d := \norm{\cdot}_{L_d(X \times \mathcal{T})}$ under the uniform measure on $X \times \mathcal{T}$, set $a_x(x', T) := \E_{y \in T}[A'_x(x', y)]$ and $b_x(x', T) := \E_{y \in Y}[A'_x(x', y)]$. By Fubini's theorem in Step~3, $U_x = \E_{x', T}[a_x(x', T)^d] = \norm{a_x}_d^d$, whereas $\norm{b_x}_d = \norm{A'_x}_{U(1, d)} = \deg_A(x)\norm{A_x}_{U(1, d)}$. For each $x' \in X$, applying~\eqref{eq:oblivious-sampler} to $y \mapsto A'_x(x', y) \in [0, 1]$ gives $\E_{T \in \mathcal{T}}[\abs{a_x(x', T) - b_x(x', T)}^d] \le \varepsilon_{\mathrm{samp}}^d + \delta_{\mathrm{samp}}$; averaging over $x' \in X$ and applying Minkowski's reverse triangle inequality yields
\[
    \abs{U_x^{1/d} - \norm{A'_x}_{U(1, d)}}
    = \abs{\norm{a_x}_d - \norm{b_x}_d}
    \le \norm{a_x - b_x}_d
    \le \bigl(\varepsilon_{\mathrm{samp}}^d + \delta_{\mathrm{samp}}\bigr)^{1/d}
    \le \varepsilon_{\mathrm{samp}} + \delta_{\mathrm{samp}}^{1/d}
    = \alpha',
\]
and dividing by $\deg_A(x)$ gives~\eqref{eq:approx-grid-norm-err}. The $O(n^{5/2}(d/\alpha')^{O(d)})$ running-time and exact-rational bit-length verification from \Cref{prop:sampler} is given in \Cref{app:sampling}.
\end{proof}

\subsection{Dyadic Sifting}\label{sec:2d-sift}

\begin{algorithm}[t]
\caption{$\Sift(X, Y, A, \varepsilon, d)$}\label{alg:sift}
\small
\noindent\textbf{Input:} Non-empty $X, Y$; matrix $A \in \{0, 1\}^{X \times Y}$ with $(1 + \varepsilon/2)\dens{A} > 2^{-d}$; $\varepsilon = 1/160$; integer $d \ge 2$.\\
\textbf{Output:} Witness tuple $(X', Y', s)$ ($s \in \{0, \dots, d-1\}$) or regularity certificate (\tagreg{}).
\begin{algsteps}[nosep, topsep=1pt]
    \item Let $\delta := (1 + \varepsilon/2)\dens{A} > 2^{-d}$, $\varepsilon_0 := \varepsilon/4$, and $\delta_s := 2^s\delta$ for $0 \le s \le d-1$.\label{line:sift-params}
    \item \textbf{for} $s = 0, 1, \dots, d-1$ \textbf{do}\algcomment{Outer dyadic row-degree check}\label{line:sift-outer-loop}
    \item \quad Let $X'_s := \{x \in X : \delta_s \le \deg_A(x) < 2\delta_s\}$; \textbf{if} $\abs{X'_s} \ge (\varepsilon/16) 2^{-(s+2)d}\abs{X}$ \textbf{then return} $(X'_s, Y, s)$.\label{line:sift-outer-ret}
    \item Let $X_{\ge \delta^2} := \{x \in X : \deg_A(x) \ge \delta^2\}$; \textbf{if} $X_{\ge \delta^2} = \emptyset$ \textbf{then return} \tagreg{}.\algcomment{Witness-row cutoff}\label{line:sift-xdelta2}
    \item Compute $\{V_x\} \gets \EstimateNorm(X, Y, A, d, \alpha')$ with $\alpha' := (\varepsilon_0 \delta^3)/8 = (\varepsilon \delta^3)/32$.\label{line:sift-est}
    \item Pick $x^* \in X_{\ge \delta^2}$ maximizing $V_{x^*}$, and let $Y' := Y_{x^*}$ and $A_{x^*} := A[X, Y']$.\label{line:sift-witness}
    \item \textbf{for} $s = 0, 1, \dots, d-1$ \textbf{do}\algcomment{Inner dyadic row-degree check}\label{line:sift-inner-loop}
    \item \quad Let $X''_s := \{x' \in X : \delta_s \le \deg_{A_{x^*}}(x') < 2\delta_s\}$; \textbf{if} $\abs{X''_s} \ge (\varepsilon/16) 2^{-(s+2)d}\abs{X}$ \textbf{then return} $(X''_s, Y', s)$.\label{line:sift-inner-ret}
    \item \textbf{return} \tagreg{}.\label{line:sift-reg}
\end{algsteps}
\end{algorithm}

\begin{lemma}[Dyadic sifting]\label{lem:dyadic-sift}
For $\varepsilon = 1/160$, integer $d \ge 2$, and $\delta := (1 + \frac{\varepsilon}{2})\dens{A} > 2^{-d}$, $\Sift(X, Y, A, \varepsilon, d)$ (\Cref{alg:sift}) runs in $O(n^{5/2} 2^{O(d^2)})$ deterministic time and satisfies:
\begin{enumerate}[nosep, leftmargin=*, topsep=2pt]
    \item If it returns $(X', Y', s)$, then $X' \subseteq X$ and $Y' \subseteq Y$ are non-empty, $0 \le s \le d-1$, $2^s\delta \le \dens{A[X', Y']} \le 1$, and
    \begin{equation}\label{eq:dyadic-sift-guarantees}
        \dens{A[X', Y']} \ge 2^s\Bigl(1 + \frac{\varepsilon}{2}\Bigr)\dens{A}, \quad
        \abs{X'} \ge \frac{\varepsilon}{16} 2^{-(s+2)d}\abs{X}, \quad
        \abs{Y'} \ge \Bigl(1 + \frac{\varepsilon}{2}\Bigr)^2\dens{A}^2 \abs{Y} > 2^{-2d}\abs{Y}.
    \end{equation}
    \item If it returns a regularity certificate ($\tagreg$), then $A$ is $(\varepsilon, 2, d)$-regular.
\end{enumerate}
\end{lemma}

\begin{proof}
For~(1), any slice returned at \Cref{line:sift-outer-ret} or \Cref{line:sift-inner-ret} has row degrees in $[\delta_s, 1]$ on $Y'$ and size $\abs{X'} \ge (\varepsilon/16)2^{-(s+2)d}\abs{X} > 0$, giving $2^s\delta \le \dens{A[X', Y']} \le 1$ (forcing $\delta \le 1$) and $\abs{Y'} \ge \delta^2\abs{Y} > 2^{-2d}\abs{Y}$.

For~(2), if $\delta > 1$ then $\norm{A}_{U(2, d)} \le 1 < (1 + \varepsilon)\dens{A}$, and \Cref{line:sift-xdelta2} returns $\tagreg$ in $O(n^2 + dn)$ time without calling $\EstimateNorm$. Assume now $\delta \le 1$ and $\Sift$ returns $\tagreg$. Since $2^d\delta > 1$, the $d$ slices $[\delta_s, 2\delta_s)$ cover $[\delta, 1]$; using $d 2^{-d} \le 1/2$ ($d \ge 2$), any $h : X \to [0, 1]$ with fewer than $(\varepsilon/16)2^{-(s+2)d}\abs{X}$ elements per slice obeys
\[
    \E_{x \in X}\bigl[\max(\delta, h(x))^d\bigr]
    \le \delta^d + \sum_{s=0}^{d-1} \frac{\varepsilon}{16} 2^{-(s+2)d}(2^{s+1}\delta)^d
    = \Bigl(1 + \frac{\varepsilon}{16} d 2^{-d}\Bigr)\delta^d
    \le \Bigl(1 + \frac{\varepsilon}{32}\Bigr)\delta^d.
\]
This holds for $h = \deg_A$ by \Cref{line:sift-outer-loop,line:sift-outer-ret}. If every $x \in X_{\ge \delta^2}$ has $\norm{A_x}_{U(1, d)} < (1 + \varepsilon_0)\delta$, then $\deg_A(x)^d\norm{A_x}_{U(1, d)}^d \le (1 + \varepsilon_0)^d\delta^d\max(\delta, \deg_A(x))^d$ for all $x \in X$; averaging via~\eqref{eq:grid-norm-fubini} and using $(1 + \varepsilon_0)^d \ge 1 + \frac{\varepsilon}{2} > 1 + \frac{\varepsilon}{32}$ gives $\norm{A}_{U(2, d)} < (1 + \varepsilon_0)\delta < (1 + \varepsilon)\dens{A}$. Otherwise, some $x_{\mathrm{wit}} \in X_{\ge \delta^2}$ has $\norm{A_{x_{\mathrm{wit}}}}_{U(1, d)} \ge (1 + \varepsilon_0)\delta$. By \Cref{lem:approx-grid-norm}, $\abs{v_x - \norm{A_x}_{U(1, d)}} \le \alpha'/\delta^2 = \varepsilon_0\delta/8$ on $X_{\ge \delta^2}$, so maximizing $V_x = v_x^d$ (\Cref{line:sift-witness}) gives $\norm{A_{x^*}}_{U(1, d)} \ge (1 + \frac{3}{4}\varepsilon_0)\delta$. Since $(1 + \frac{3}{4}\varepsilon_0)^d \ge 1 + \frac{3}{8}\varepsilon > 1 + \frac{\varepsilon}{32}$, the slice bound on $h = \deg_{A_{x^*}}$ would fail at \Cref{line:sift-inner-loop,line:sift-inner-ret}, a contradiction. Finally, $\alpha' \in (\frac{\varepsilon}{32}2^{-3d}, \frac{\varepsilon}{32}] \subset (0, 1)$ gives time $O(n^{5/2} 2^{O(d^2)})$ by \Cref{lem:approx-grid-norm}.
\end{proof}

\subsection{Capped Rectangle Decomposition and Riemann-Sum Telescoping}\label{sec:2d-decomp}

\begin{algorithm}[t]
\caption{$\FindRect(X, Y, A, \varepsilon, d, \alpha_{\mathrm{cap}})$ and $\DecomposeTwoD(X, Y, A, \varepsilon, d, \alpha_{\mathrm{low}}, \alpha_{\mathrm{cap}})$}\label{alg:2d}
\small
\noindent\textbf{Procedure $\FindRect(X, Y, A, \varepsilon, d, \alpha_{\mathrm{cap}})$} \quad (\textbf{Requires:} $\dens{A} \ge 2^{-d}$, $d \ge 2$, $\alpha_{\mathrm{cap}} > 0$)
\begin{algsteps}[nosep, topsep=1pt]
    \item \textbf{if} $\dens{A} \ge \alpha_{\mathrm{cap}}$ \textbf{then return} $(X, Y, \tagdenseA)$.\label{line:findrect-cap1}
    \item Let $(X', \mathrm{status}) \gets \Prune(X, Y, A, \varepsilon, 1/2)$ and $A' := A[X', Y]$.\label{line:findrect-prune}
    \item \textbf{if} $\mathrm{status} = \textnormal{\emph{boosted}}$ or $\dens{A'} \ge (1 + \varepsilon/2)\dens{A}$ \textbf{then return} $\FindRect(X', Y, A', \varepsilon, d, \alpha_{\mathrm{cap}})$.\label{line:findrect-rec1}
    \item \textbf{if} $\dens{A'} \ge \alpha_{\mathrm{cap}}$ \textbf{then return} $(X', Y, \tagdenseA)$.\label{line:findrect-cap2}
    \item \textbf{if} $\Sift(X', Y, A', \varepsilon, d)$ returns $(X'', Y'', s)$ \textbf{then return} $\FindRect(X'', Y'', A[X'', Y''], \varepsilon, d, \alpha_{\mathrm{cap}})$.\label{line:findrect-sift}
    \item \textbf{return} $(X', Y, \tagreg)$.\label{line:findrect-reg}
\end{algsteps}
\vspace{1pt}
\noindent\textbf{Procedure $\DecomposeTwoD(X, Y, A, \varepsilon, d, \alpha_{\mathrm{low}}, \alpha_{\mathrm{cap}})$} \quad (\textbf{Requires:} $d \ge 2$, $0 < \alpha_{\mathrm{low}} < \alpha_{\mathrm{cap}}$)
\begin{algsteps}[resume, nosep, topsep=1pt]
    \item Initialize $\ell \gets 1$, $A^{(1)} \gets A$, and $\tau_{\mathrm{low}} := \max(2^{-d}, \alpha_{\mathrm{low}})$.\label{line:dec2d-init}
    \item \textbf{while} $\dens{A^{(\ell)}} > \tau_{\mathrm{low}}$ \textbf{do}\label{line:dec2d-while}
    \item \quad Let $(X_\ell, Y_\ell, \mathrm{status}_\ell) \gets \FindRect(X, Y, A^{(\ell)}, \varepsilon, d, \alpha_{\mathrm{cap}})$ and $A_\ell := A^{(\ell)}[X_\ell, Y_\ell]$.\label{line:dec2d-call}
    \item \quad Set $A^{(\ell+1)} \gets A^{(\ell)} - A_\ell$ and $\ell \gets \ell + 1$.\label{line:dec2d-sub}
    \item Set $L \gets \ell$, $(X_L, Y_L, A_L, \mathrm{status}_L) \gets (X, Y, A^{(L)}, \tagsparse)$, and \textbf{return} $\{(X_\ell, Y_\ell, A_\ell, \mathrm{status}_\ell)\}_{\ell=1}^L$.\label{line:dec2d-ret}
\end{algsteps}
\end{algorithm}

\begin{lemma}[Rectangle search]\label{lem:find-rect}
Fix $\varepsilon = 1/160$, integer $d \ge 2$, and $0 < \alpha_{\mathrm{low}} < \alpha_{\mathrm{cap}}$, and set $\tau_{\mathrm{low}} := \max(2^{-d}, \alpha_{\mathrm{low}})$ and $W := \alpha_{\mathrm{cap}}/\tau_{\mathrm{low}}$. Whenever $\dens{A^{(\ell)}} > \tau_{\mathrm{low}}$, $\FindRect(X, Y, A^{(\ell)},\allowbreak \varepsilon, d, \alpha_{\mathrm{cap}})$ (\Cref{alg:2d}) returns non-empty $X_\ell \subseteq X, Y_\ell \subseteq Y$ after $r \le \min\bigl(2d/\varepsilon,\, 1 + \frac{2\log_2\max(1, W)}{\varepsilon}\bigr)$ recursive steps with non-negative integer scales $s_1, \dots, s_r$ summing to $S_\ell := \sum_{t=1}^r s_t \in \{0, \dots, d\}$ such that, for $A_\ell := A^{(\ell)}[X_\ell, Y_\ell]$,
\begin{equation}\label{eq:capped-goodrect-shrinkage}
    \dens{A_\ell} \ge 2^{S_\ell}\dens{A^{(\ell)}}, \qquad
    \abs{X_\ell} \ge \frac{1}{3}\left(\frac{\varepsilon}{48} 2^{-2d}\right)^r 2^{-S_\ell d}\abs{X}, \qquad
    \abs{Y_\ell} \ge 2^{-2dr}\abs{Y}.
\end{equation}
A regular return ($\mathrm{status}_\ell = \tagreg$) is $(\varepsilon, 2, d)$-regular and $\varepsilon$-min-degree with $\dens{A^{(\ell)}} \le \dens{A_\ell} < \alpha_{\mathrm{cap}}$; a density-boosted return ($\mathrm{status}_\ell = \tagdenseA$) satisfies $\dens{A_\ell} \ge \max(\alpha_{\mathrm{cap}}, 2^{S_\ell}\dens{A^{(\ell)}})$.
\end{lemma}

\begin{proof}
Track $(X_t, Y_t, A_t)$ from $(X_0, Y_0, A_0) := (X, Y, A^{(\ell)})$. When $r = 0$, halting at \Cref{line:findrect-cap1} (forced if $W \le 1$) or after one certified prune (\Cref{line:findrect-prune,line:findrect-cap2,line:findrect-reg}) gives $S_\ell = 0$, $\dens{A_\ell} \ge \dens{A_0}$, $\abs{X_\ell} \ge \frac{1}{3}\abs{X}$, and $Y_\ell = Y$ by \Cref{lem:min-degree}(1). When $r \ge 1$, $\tau_{\mathrm{low}} < \dens{A_0} \le \dens{A_{r-1}} < \alpha_{\mathrm{cap}}$ (so $W > 1$). Assigning $s_t := 0$ to pruning recursions (\Cref{line:findrect-rec1}), \Cref{lem:min-degree,lem:dyadic-sift} give $\dens{A_t} \ge 2^{s_t}(1 + \varepsilon/2)\dens{A_{t-1}}$, $\abs{X_t} \ge \frac{\varepsilon}{48}2^{-(s_t+2)d}\abs{X_{t-1}}$, and $\abs{Y_t} \ge 2^{-2d}\abs{Y_{t-1}}$ for each $t \in [r]$. At most one final certified prune follows step $r$, proving~\eqref{eq:capped-goodrect-shrinkage} and $1 \ge \dens{A_\ell} \ge 2^{S_\ell}(1 + \varepsilon/2)^r \dens{A_0}$. Using $\log_2(1 + \varepsilon/2) \ge \varepsilon/2$ gives $S_\ell + \frac{\varepsilon}{2}r \le \log_2(1/\dens{A_0}) < d$, so $S_\ell \le d$ and $r \le 2d/\varepsilon$; similarly, $(1 + \varepsilon/2)^{r-1} < W$ gives $r \le 1 + \frac{2\log_2 W}{\varepsilon}$.
\end{proof}

\begin{theorem}[Capped 2D decomposition]\label{thm:2d-decomp}
Fix $\varepsilon = 1/160$, integer $d \ge 2$, $(\alpha_{\mathrm{low}}, \alpha_{\mathrm{cap}}) = (2^{-40}\alpha, 2^{40}\alpha)$ ($\alpha > 0$), and $M_{\mathrm{rect}} := 3\bigl(\frac{48}{\varepsilon} 2^{2d}\bigr)^{161/\varepsilon} 2^{322d/\varepsilon} = 2^{O(d)}$. Then $\DecomposeTwoD(X, Y,\allowbreak A,\allowbreak \varepsilon,\allowbreak d,\allowbreak \alpha_{\mathrm{low}}, \alpha_{\mathrm{cap}})$ (\Cref{alg:2d}) runs in $O(n^{5/2} 2^{O(d^2)})$ deterministic time and decomposes $A = \sum_{\ell=1}^L A_\ell$ into $L \le 1 + 2^d M_{\mathrm{rect}} 2^{d^2} = 2^{O(d^2)}$ binary rectangles $A_\ell \in \{0, 1\}^{X_\ell \times Y_\ell}$ on non-empty $X_\ell \subseteq X, Y_\ell \subseteq Y$ with pairwise disjoint supports:
\begin{enumerate}[nosep, leftmargin=*, topsep=2pt]
    \item Every $\ell \in [L-1]$ obeys $\abs{X_\ell}\abs{Y_\ell} \ge M_{\mathrm{rect}}^{-1} 2^{-S_\ell d}\abs{X}\abs{Y}$ and $\abs{Y_\ell} \ge 2^{-322d/\varepsilon}\abs{Y}$; for each $s \in \{0, \dots, d\}$, at most $L_s \le 2^d M_{\mathrm{rect}} 2^{sd}$ rectangles $\ell \in [L-1]$ have $S_\ell = s$.
    \item Every regular rectangle $\ell \in [L-1]$ (tagged $\tagreg$) is $(\varepsilon, 2, d)$-regular and $\varepsilon$-min-degree with $\tau_{\mathrm{low}} < \dens{A_\ell} < 2^{40}\alpha$ and $S_\ell \le 79$, and their count satisfies $L_{\mathrm{reg}} \le \sum_{s=0}^{\min(79, d)} L_s \le 80 \cdot 2^{80d} M_{\mathrm{rect}} = 2^{O(d)}$.
    \item Every density-boosted rectangle $\ell \in [L-1]$ (tagged $\tagdenseA$) obeys $\dens{A_\ell} \ge \max(2^{40}\alpha, 2^{S_\ell - 40}\alpha)$, with $\sum_{\ell \in [L-1] : \mathrm{status}_\ell = \tagdenseA} \dens{A_\ell}\abs{X_\ell}\abs{Y_\ell} \le \dens{A}\abs{X}\abs{Y}$.
    \item The sparse residual $\ell = L$ (tagged $\tagsparse$) has $(X_L, Y_L) = (X, Y)$ and $\dens{A_L} \le \tau_{\mathrm{low}} = \max(2^{-d}, 2^{-40}\alpha)$.
\end{enumerate}
\end{theorem}

\begin{proof}
Zeroing out $X_\ell \times Y_\ell$ at \Cref{line:dec2d-sub} preserves binarity and disjoint supports. For $\ell \in [L-1]$, $W \le 2^{80}$ gives $r \le 1 + 160/\varepsilon \le 161/\varepsilon$ in \Cref{lem:find-rect}, proving~(1)'s size bounds; since each piece with $S_\ell = s$ has $\dens{A_\ell} > 2^{-d}$, it drops $\dens{A^{(\ell)}}$ by $> 2^{-d} M_{\mathrm{rect}}^{-1} 2^{-sd}$ from $\dens{A^{(1)}} \le 1$, bounding $L_s$ and $L$. Regular pieces satisfy $2^{40}\alpha > \dens{A_\ell} > 2^{S_\ell - 40}\alpha$, forcing $S_\ell \le 79$ and~(2); disjoint supports and loop exit give~(3)--(4), and $(L - 1)(1 + 161/\varepsilon)$ calls to $\Prune$ and $\Sift$ take $O(n^{5/2} 2^{O(d^2)})$ time.
\end{proof}

\begin{lemma}[Riemann-sum telescoping; proof in \Cref{app:helper-proofs}]\label{lem:riemann}
For integer $d \ge 2$, any $0 < \alpha_{\mathrm{low}} < \alpha_{\mathrm{cap}}$, and every $p \in (0, 1)$, the output $\{(X_\ell, Y_\ell, A_\ell, \mathrm{status}_\ell)\}_{\ell=1}^L$ of $\DecomposeTwoD$ satisfies
\begin{equation}\label{eq:riemann-telescoping}
    \sum_{\ell=1}^L \dens{A_\ell}^p \abs{X_\ell}\abs{Y_\ell} \le \frac{1}{p}\dens{A}^p \abs{X}\abs{Y} \qquad \left(\textrm{in particular, } \le 4\dens{A}^{1/4}\abs{X}\abs{Y} \textrm{ for } p = \frac{1}{4}\right).
\end{equation}
\end{lemma}

\section{Three-Dimensional Regularity Decomposition}\label{sec:3d}

Throughout this section, $X, Y, Z$ are finite non-empty sets ($n := \abs{X} + \abs{Y} + \abs{Z}$),
$A \in \{0,1\}^{X \times Y}$ and $B \in \{0,1\}^{Y \times Z}$ are binary matrices, $d \ge 640$ is an even
integer, and we fix
\[
    \varepsilon := \frac{1}{160}, \qquad
    \gamma := 2^{-12} = \frac{1}{4096}, \qquad
    H := \lceil 3d/4 \rceil, \qquad
    \Lambda_h := 1024 \cdot 2^h \quad (0 \le h \le H).
\]

\subsection{Dyadic Column-Degree Bucketing}\label{sec:3d-bucketing}

To keep $A[X, Y^*]$ bounded when $\FindCube$ restricts $Y$ to $Y^* \subseteq Y$, we maintain the column-degree
invariant $\max_{y \in Y}\deg_{A^T}(y) \le 2\alpha$, and restore it on row-restricted children via $d$ dyadic
column-degree buckets:

\begin{lemma}[Dyadic column-degree bucketing; proof in \Cref{app:helper-proofs}]\label{lem:dyadic-bucketing}
Let $A' \in \{0,1\}^{X' \times Y'}$ and $B' \in \{0,1\}^{Y' \times Z'}$ be binary matrices on finite
non-empty sets $X', Y', Z'$, with $\dens{A'} \le \mu$ for some $\mu \in (2^{-d}, 1]$. Partition $Y'$ into
\begin{equation}\label{eq:dyadic-buckets}
    Y'_0 := \big\{y \in Y' : \deg_{(A')^T}(y) \le 2\mu\big\}, \qquad
    Y'_j := \big\{y \in Y' : 2^j\mu < \deg_{(A')^T}(y) \le 2^{j+1}\mu\big\} \quad (1 \le j < d).
\end{equation}
Then $Y' = \bigsqcup_{j=0}^{d-1} Y'_j$. For each non-empty bucket $Y'_j$, setting
$\alpha'_j := 2^j\mu \in (0, 1]$ gives $\max_{y \in Y'_j}\deg_{A'[X', Y'_j]^T}(y) \le 2\alpha'_j$, and
\[
    \sum_{j \in \{0, \dots, d-1\} : Y'_j \ne \emptyset} (\alpha'_j)^{1/4}\dens{B'[Y'_j, Z']}^{1/4}\abs{Y'_j}
    \le \frac{7}{2}\,\mu^{1/4}\dens{B'}^{1/4}\abs{Y'}.
\]
\end{lemma}

\subsection{Cube Search and the 3D Decomposition Algorithm}\label{sec:3d-alg}

\Cref{alg:3d} gives $\FindCube$ and $\DecomposeThreeD$ ($\cup$ concatenates lists; submatrices are
zero-extended as in \Cref{sec:notation}, and $\lfloor \frac{1}{2}\log_2(p/q) \rfloor$ on rationals $p/q \ge 1$
is evaluated exactly via $\max\{k \in \N : 4^k q \le p\}$).

\begin{algorithm}[t]
\caption{$\FindCube(X, Y, Z, A, B, \alpha, \beta_{\mathrm{init}})$ and $\DecomposeThreeD(X, Y, Z, A, B, \alpha, h)$}\label{alg:3d}
\footnotesize
\noindent\textbf{Procedure $\FindCube(X, Y, Z, A, B, \alpha, \beta_{\mathrm{init}})$:} \quad (\textbf{Requires:} $\dens{B} \ge \beta_{\mathrm{init}} > 2^{-d}$ and $\alpha > 0$)
\begin{algsteps}[nosep,topsep=1pt]
    \item \textbf{if} $\dens{B} \ge 2^{40}\beta_{\mathrm{init}}$ \textbf{then return} $(\tagdenseB, Y, Z, \emptyset)$.\algcomment{Cash $B$-density jump}\label{line:cube-cap}
    \item Let $(Y', \mathrm{status}_1) \gets \Prune(Y, Z, B, \varepsilon\gamma/2, 1/2)$ and $B' := B[Y', Z]$.\algcomment{Row density floor}\label{line:cube-prune1}
    \item \textbf{if} $\mathrm{status}_1 = \textnormal{\emph{boosted}}$ or $\dens{B'} \ge (1 + \frac{\varepsilon\gamma}{4})\dens{B}$ \textbf{then return} $\FindCube(X, Y', Z, A[X, Y'], B', \alpha, \beta_{\mathrm{init}})$.\label{line:cube-rec1}
    \item Let $\{(X_\ell, Y_\ell, A_\ell, \mathrm{status}_\ell)\}_{\ell=1}^L \gets \DecomposeTwoD(X, Y', A[X, Y'], \varepsilon, d, 2^{-40}\alpha, 2^{40}\alpha)$.\label{line:cube-dec2d}
    \item Set $Z_L \gets Z$ if $\dens{A_L} \le 2^{-d}$, and $Z_L \gets \emptyset$ otherwise.\algcomment{Test 3D sparsity}\label{line:cube-zl}
    \item \textbf{for} $\ell = 1, \dots, L-1$ \textbf{do}\label{line:cube-loop}
    \item \quad \textbf{if} $\dens{B[Y_\ell, Z]} \ge 2\dens{B'}$ \textbf{then return} $\FindCube(X, Y_\ell, Z, A[X, Y_\ell], B[Y_\ell, Z], \alpha, \beta_{\mathrm{init}})$.\label{line:cube-slice}
    \item \quad \textbf{if} $\mathrm{status}_\ell = \tagdenseA$ \textbf{then} set $Z_\ell \gets \emptyset$ and \textbf{continue}.\algcomment{Defer boosted $A$}\label{line:cube-densea}
    \item \quad Let $(Z_\ell, \mathrm{status}_2) \gets \Prune(Z, Y_\ell, B^T[Z, Y_\ell], \varepsilon, \gamma)$ and $B_\ell := B[Y_\ell, Z_\ell]$.\label{line:cube-prune2}
    \item \quad \textbf{if} $\mathrm{status}_2 = \textnormal{\emph{boosted}}$ or $\dens{B_\ell} \ge (1 + \varepsilon\gamma)\dens{B[Y_\ell, Z]}$ \textbf{then return} $\FindCube(X, Y_\ell, Z_\ell, A[X, Y_\ell], B_\ell, \alpha, \beta_{\mathrm{init}})$.\label{line:cube-rec2}
    \item \quad \textbf{if} $\Sift(Z_\ell, Y_\ell, B_\ell^T, \varepsilon, d) = (Z'', Y'', s)$ \textbf{then return} $\FindCube(X, Y'', Z'', A[X, Y''], B[Y'', Z''], \alpha, \beta_{\mathrm{init}})$.\label{line:cube-sift}
    \item \textbf{return} $(\tagnormal, Y', Z, \{(X_\ell, Y_\ell, Z_\ell, A_\ell, \mathrm{status}_\ell)\}_{\ell=1}^L)$.\algcomment{Aligned cube}\label{line:cube-ret}
\end{algsteps}
\vspace{1pt}
\noindent\textbf{Procedure $\DecomposeThreeD(X, Y, Z, A, B, \alpha, h)$:} \quad (\textbf{Requires:} $\alpha \in (0,1]$, $\max_{y \in Y}\deg_{A^T}(y) \le 2\alpha$, $0 \le h \le H$)
\begin{algsteps}[resume,nosep,topsep=1pt]
    \item \textbf{if} $\dens{A} \le 2^{-d}$ or $\dens{B} \le 2^{-d}$ or $\abs{Y}\abs{Z} < 2^d$ \textbf{then return} $\{(X, Y, Z, A, B)\}$.\algcomment{Sparse base case}\label{line:3d-base1}
    \item \textbf{if} $h = H$ \textbf{then} split $\supp(B)$ into $N_{\mathrm{split}} := \lceil \nnz(B)/m \rceil$ batches $B_s$ of size $\le m := \lfloor 2^{-d}\abs{Y}\abs{Z} \rfloor$ and \textbf{return} $\{(X, Y, Z, A, B_s)\}_{s=1}^{N_{\mathrm{split}}}$.\label{line:3d-base2}
    \item Let $(\mathrm{status}, Y^*, Z^*, \mathcal{L}) \gets \FindCube(X, Y, Z, A, B, \alpha, \dens{B})$.\label{line:3d-findcube}
    \item Let $\mathcal{S}^* \gets \DecomposeThreeD(X, Y, Z, A, B - B[Y^*, Z^*], \alpha, h)$.\algcomment{Same-depth residual}\label{line:3d-sstar}
    \item \textbf{if} $\mathrm{status} = \tagdenseB$ \textbf{then}\label{line:3d-brancha}
    \item \quad Let $\rho_B := \dens{B[Y^*, Z^*]}/\dens{B}$ and $\Delta h_B := \min\bigl(H - h,\, \lfloor \frac{1}{2}\log_2 \rho_B \rfloor\bigr)$.\label{line:3d-dhb}
    \item \quad \textbf{return} $\mathcal{S}^* \cup \DecomposeThreeD(X, Y^*, Z^*, A[X, Y^*], B[Y^*, Z^*], \alpha, h + \Delta h_B)$.\label{line:3d-recb}
    \item Unpack $\mathcal{L} = \{(X_\ell, Y_\ell, Z_\ell, A_\ell, \mathrm{status}_\ell)\}_{\ell=1}^L$, put $R_\ell := Z^* \setminus Z_\ell$, and $\mathcal{S}_{\mathrm{direct}} := \{(X_\ell, Y_\ell, Z_\ell, A_\ell, B[Y_\ell, Z_\ell]) : Z_\ell \ne \emptyset\}$.\label{line:3d-direct}
    \item \textbf{for each} $\ell \in [L]$ with $R_\ell \ne \emptyset$ \textbf{do}\label{line:3d-loop}
    \item \quad Let $\mu_\ell := \dens{A_\ell}$; set $\Delta h_\ell := \min\bigl(H - h,\, \lfloor \frac{1}{2}\log_2(\mu_\ell/\alpha) \rfloor\bigr)$ if $\mathrm{status}_\ell = \tagdenseA$, and $\Delta h_\ell := 1$ otherwise.\label{line:3d-dhl}
    \item \quad Partition $Y_\ell = \bigsqcup_{j=0}^{d-1} Y_{\ell,j}$ via~\eqref{eq:dyadic-buckets} at $\mu = \mu_\ell$, and set $\mathcal{S}_\ell := \bigcup_{j : Y_{\ell,j} \ne \emptyset} \DecomposeThreeD(X_\ell, Y_{\ell,j}, R_\ell,\allowbreak A_\ell[X_\ell, Y_{\ell,j}],\allowbreak B[Y_{\ell,j}, R_\ell], 2^j\mu_\ell, h + \Delta h_\ell)$.\label{line:3d-reca}
    \item \textbf{return} $\mathcal{S}_{\mathrm{direct}} \cup \mathcal{S}^* \cup \bigcup_{\ell \in [L] : R_\ell \ne \emptyset} \mathcal{S}_\ell$.\label{line:3d-ret}
\end{algsteps}
\end{algorithm}

\begin{lemma}[Single cube-extraction step; proof in \Cref{app:helper-proofs}]\label{lem:cube-step}
For $\alpha > 0$ and $\beta_{\mathrm{init}} = \dens{B} > 2^{-d}$, invoking
$\FindCube(X, Y, Z,\allowbreak A, B, \alpha, \beta_{\mathrm{init}})$ makes
$r \le R_{\mathrm{cube}} := 1 + \lfloor 160/(\varepsilon\gamma) \rfloor = O(1)$ restarts and returns
$(\mathrm{status}, Y^*, Z^*, \mathcal{L})$ in deterministic time $O(n^{5/2} 2^{O(d^2)})$. Putting
$s_t = 0$ on non-sifting restarts and $s_t = s$ on sifting restarts, $S_B := \sum_{t=1}^r s_t \in \{0, \dots, d\}$
and $\beta^* := \dens{B[Y^*, Z^*]}$ satisfy $\beta^* \ge 2^{S_B}\beta_{\mathrm{init}} > 2^{S_B - d}$ and
$\abs{Y^*}\abs{Z^*} \ge M_{\mathrm{cube}}^{-1} 2^{-S_B d}\abs{Y}\abs{Z}$ for $M_{\mathrm{cube}} = 2^{O(d)}$.
A $B$-boosted return ($\mathrm{status} = \tagdenseB$) obeys $\beta^* \ge \max(2^{40}, 2^{S_B})\beta_{\mathrm{init}}$;
a normal return ($\mathrm{status} = \tagnormal$) obeys
$\beta^* < 2^{40}(1 + \frac{\varepsilon\gamma}{4})\beta_{\mathrm{init}} < 2^{41}\beta_{\mathrm{init}}$,
$S_B \le 40$, $(X_L, Y_L) = (X, Y^*)$, and for each $\ell \in [L-1]$ of scale $S_\ell$:
\begin{enumerate}[nosep,leftmargin=*,topsep=2pt]
    \item On a regular slice ($\mathrm{status}_\ell = \tagreg$), $Z_\ell \ne \emptyset$, both $A_\ell$
    and $B_\ell^T := B^T[Z_\ell, Y_\ell]$ are $(\varepsilon, 2, d)$-regular and $\varepsilon$-min-degree,
    $\max(2^{-d}, 2^{-40}\alpha) < \dens{A_\ell} < 2^{40}\alpha$, $S_\ell \le 79$,
    $(1 - \frac{\varepsilon\gamma}{2})\beta^* \le \dens{B_\ell} \le 3\beta^*$, and
    $\abs{Z^* \setminus Z_\ell} < \gamma\abs{Z^*}$; if $Z^* \setminus Z_\ell \ne \emptyset$, then
    $\dens{B[Y_\ell, Z^* \setminus Z_\ell]} \le 2\beta^*$.
    \item On an $A$-boosted slice ($\mathrm{status}_\ell = \tagdenseA$),
    $\dens{A_\ell} \ge \max(2^{40}\alpha, 2^{S_\ell - 40}\alpha)$, $Z_\ell = \emptyset$, and
    $\dens{B[Y_\ell, Z^*]} < 2\beta^*$.
\end{enumerate}
\end{lemma}

\subsection{Structural Invariants and Exact Decomposition}\label{sec:3d-invariants}

\begin{lemma}[Recursive invariants and exact decomposition; proof in \Cref{app:helper-proofs}]\label{lem:children-invariants}
Suppose we call $\DecomposeThreeD$ on $(X, Y, Z,\allowbreak A, B, \alpha, h)$ for finite non-empty $X, Y, Z$
with $\alpha \in (0, 1]$, $h \in \{0, \dots, H\}$, and $\max_{y \in Y}\deg_{A^T}(y) \le 2\alpha$:
\begin{enumerate}[nosep,leftmargin=*,topsep=2pt]
    \item Every recursive child satisfies these preconditions and strictly decreases $(H - h, \nnz(B))$ lexicographically.
    \item The call terminates and returns $\mathcal{S} = \{(X_k, Y_k, Z_k, A_k, B_k)\}_{k=1}^K$ with
    $0 \le A_k \le A[X_k, Y_k]$, $0 \le B_k \le B[Y_k, Z_k]$, and $AB = \sum_{k=1}^K A_k B_k$.
    \item Every returned piece is sparse or regular (\Cref{def:pieces}).
\end{enumerate}
\end{lemma}

\subsection{Fractional Potential Analysis}\label{sec:3d-potential}

For an interior call at depth $h < H$, set $\beta := \dens{B}$, $\beta^* := \dens{B[Y^*, Z^*]} \ge \beta$,
$q := \abs{Y^*}\abs{Z^*}/(\abs{Y}\abs{Z}) \in (0, 1]$, and
$W^* := \Lambda_h \alpha^{1/4}(\beta^*)^{1/4}\abs{X}\abs{Y^*}\abs{Z^*}$. Since $B - B[Y^*, Z^*]$ has
density $\beta_{\mathrm{res}} = \beta(1 - q\beta^*/\beta)$ with $q\beta^*/\beta \in (0, 1]$, concavity
$(1 - u)^{1/4} \le 1 - \frac{1}{4}u$ gives
\begin{equation}\label{eq:concavity-drop}
    \Lambda_h \alpha^{1/4}\beta_{\mathrm{res}}^{1/4}\abs{X}\abs{Y}\abs{Z}
    \le \Lambda_h \alpha^{1/4}\beta^{1/4}\abs{X}\abs{Y}\abs{Z}
       - \frac{1}{4}\left(\frac{\beta^*}{\beta}\right)^{3/4} W^*.
\end{equation}
This credit pays for all extracted-cube terms; \Cref{tab:potential-budget} summarizes the normal-branch
budget, where every group costs less than $\frac{1}{64}W^*$.

\begin{table}[t]
\centering\footnotesize
\setlength{\tabcolsep}{4pt}
\begin{tabularx}{\textwidth}{@{}>{\raggedright\arraybackslash}p{0.27\textwidth}>{\raggedright\arraybackslash}Xr@{}}
\toprule
\textbf{Normal-branch group} & \textbf{Saving mechanism} & \textbf{Share of $W^*$} \\
\midrule
1. Direct pieces $\mathcal{S}_{\mathrm{direct}}$ &
Riemann sum (\Cref{lem:riemann}) divided by $\Lambda_h \ge 1024$ &
$< +\frac{8}{1024} < +\frac{1}{64}$ \\
2. Remainder $\mathcal{S}_L$ ($\ell = L$) &
Window floor $\dens{A_L} \le 2^{-40}\alpha$, giving $(2^{-40})^{1/4} = 2^{-10}$ &
$+\frac{7}{1024} < +\frac{1}{64}$ \\
3. $A$-boosted children ($\tagdenseA$) &
Jump $2^{\Delta h_\ell}\mu_\ell^{1/4} \le 2^{-10}\alpha^{-3/4}\mu_\ell$ and $A$-edge conservation &
$< +\frac{14}{1024} < +\frac{1}{64}$ \\
4. Regular complements ($\tagreg$) &
Pruned column fraction $\abs{Z^* \setminus Z_\ell} < \gamma\abs{Z^*}$ ($\gamma = 2^{-12}$) &
$< +\frac{56}{4096} < +\frac{1}{64}$ \\
\midrule
\textbf{Subtotal of Rows 1--4} &
\textbf{Less than one quarter of the residual credit} &
$\mathbf{< +\frac{4}{64} = +\frac{1}{16}}$ \\
\midrule
5. Residual call $\mathcal{S}^*$ (credit) &
Concavity $(1 - u)^{1/4} \le 1 - \frac{1}{4}u$ in~\eqref{eq:concavity-drop} with $\beta^* \ge \beta$ &
$-\frac{1}{4}$ \\
\bottomrule
\end{tabularx}
\caption{Potential budget in the normal branch ($\mathrm{status} = \tagnormal$), measured in units of
$W^* = \Lambda_h \alpha^{1/4}(\beta^*)^{1/4}\allowbreak\abs{X}\abs{Y^*}\abs{Z^*}$ (net change $< -\frac{3}{16}W^*$).}
\label{tab:potential-budget}
\end{table}

\begin{lemma}[Inductive potential bound]\label{lem:potential}
Under the preconditions of \Cref{lem:children-invariants} with $\Lambda_h = 1024 \cdot 2^h$, the output
$\mathcal{S}$ of $\DecomposeThreeD(X, Y, Z, A, B, \alpha, h)$ satisfies
$\pot(\mathcal{S}) \le \Lambda_h \alpha^{1/4}\dens{B}^{1/4}\abs{X}\abs{Y}\abs{Z}$.
\end{lemma}

\begin{proof}
We induct on $(H - h, \nnz(B))$, putting $V := \abs{X}\abs{Y}\abs{Z}$ and $\beta := \dens{B}$, with
$\dens{A} \le 2\alpha$.

\smallskip\noindent\textbf{Base cases.}
In \Cref{line:3d-base1}, the sole piece costs $\le (2\alpha)^{1/4}\beta^{1/4}V \le \Lambda_h\alpha^{1/4}\beta^{1/4}V$
as $2^{1/4} < 1024 \le \Lambda_h$. At $h = H$ (\Cref{line:3d-base2}), $\beta > 2^{-d}$ and
$\tau := 2^{-d}\abs{Y}\abs{Z} \ge 1$, so $m = \lfloor \tau \rfloor > \tau/2$ ($\lfloor \tau \rfloor = 1 > \tau/2$
for $\tau < 2$, and $\lfloor \tau \rfloor > \tau - 1 \ge \tau/2$ for $\tau \ge 2$). Thus
$N_{\mathrm{split}} = \lceil \nnz(B)/m \rceil < 2 \cdot 2^d\beta + 1 < 3 \cdot 2^d\beta$ as $2^d\beta > 1$.
By H\"older's inequality ($\sum_s \dens{B_s}^{1/4} \le N_{\mathrm{split}}^{3/4}\beta^{1/4}$), using
$2^{1/4} < 2$, $3^{3/4} < 3$, $3d/4 \le H$, and $\beta \le \beta^{1/4}$, we get
$\pot(\mathcal{S}) \le (2\alpha)^{1/4}(3 \cdot 2^d\beta)^{3/4}\beta^{1/4}V < 6 \cdot 2^H \alpha^{1/4}\beta^{1/4}V < \Lambda_H\alpha^{1/4}\beta^{1/4}V$.

\smallskip\noindent\textbf{The $B$-boosted branch ($\mathrm{status} = \tagdenseB$).}
Here $\rho_B = \beta^*/\beta \ge 2^{40}$ and $\Delta h_B \le \frac{1}{2}\log_2 \rho_B$, giving
$2^{\Delta h_B} \le \rho_B^{1/2} = \rho_B^{-1/4}\rho_B^{3/4} \le 2^{-10}\rho_B^{3/4} < \frac{1}{4}\rho_B^{3/4}$.
By induction, the boosted child costs at most $2^{\Delta h_B}W^*$, paid with a $256\times$ margin
by~\eqref{eq:concavity-drop}.

\smallskip\noindent\textbf{The normal branch ($\mathrm{status} = \tagnormal$).}
Put $\mu_\ell := \dens{A_\ell}$ and $R_\ell := Z^* \setminus Z_\ell$. Since $\dens{A[X, Y^*]} \le 2\alpha$,
\Cref{lem:riemann} gives $\sum_{\ell=1}^L \mu_\ell^{1/4}\abs{X_\ell}\abs{Y_\ell} \le 4(2\alpha)^{1/4}\abs{X}\abs{Y^*}$,
and induction with \Cref{lem:dyadic-bucketing} on \Cref{line:3d-reca} gives, for $R_\ell \ne \emptyset$,
\begin{equation}\label{eq:bucketed-rec-bound}
    \pot(\mathcal{S}_\ell)
    \le \frac{7}{2}\Lambda_{h+\Delta h_\ell}\mu_\ell^{1/4}
        \dens{B[Y_\ell, R_\ell]}^{1/4}\abs{X_\ell}\abs{Y_\ell}\abs{R_\ell}.
\end{equation}
Using only $6^{1/4} < 2$, $2^{1/4} < 2$, and $\sqrt{2} < 2$, we bound the four groups in \Cref{tab:potential-budget}:
\begin{enumerate}[label=\arabic*.,nosep,leftmargin=*,topsep=2pt]
    \item \emph{Direct pieces:} Each $\ell \in [L]$ with $Z_\ell \ne \emptyset$ has $\abs{Z_\ell} \le \abs{Z^*}$
    and $\dens{B[Y_\ell, Z_\ell]} \le 3\beta^*$ (\Cref{lem:cube-step}(1), and $\beta^*$ at $\ell = L$), so
    $\pot(\mathcal{S}_{\mathrm{direct}}) \le (3\beta^*)^{1/4}\abs{Z^*} \cdot 4(2\alpha)^{1/4}\abs{X}\abs{Y^*}
    \le \frac{4 \cdot 6^{1/4}}{1024}W^* < \frac{8}{1024}W^* < \frac{1}{64}W^*$.
    \item \emph{Remainder $\mathcal{S}_L$:} When $R_L \ne \emptyset$, $(X_L, Y_L, R_L) = (X, Y^*, Z^*)$,
    $\Delta h_L = 1$, and $2^{-d} < \mu_L \le \max(2^{-d}, 2^{-40}\alpha) = 2^{-40}\alpha$, so~\eqref{eq:bucketed-rec-bound}
    gives $\pot(\mathcal{S}_L) \le \frac{7}{2}(2\Lambda_h)(2^{-40}\alpha)^{1/4}(\beta^*)^{1/4}\abs{X}\abs{Y^*}\abs{Z^*}
    = \frac{7}{1024}W^* < \frac{1}{64}W^*$.
    \item \emph{$A$-boosted children ($\mathrm{status}_\ell = \tagdenseA$):} Here $\rho_\ell := \mu_\ell/\alpha \ge 2^{40}$,
    $R_\ell = Z^*$, $\dens{B[Y_\ell, Z^*]} < 2\beta^*$, and
    $2^{\Delta h_\ell}\mu_\ell^{1/4} \le \rho_\ell^{1/2}\mu_\ell^{1/4} = \rho_\ell^{-1/4}\alpha^{-3/4}\mu_\ell \le 2^{-10}\alpha^{-3/4}\mu_\ell$.
    Using $\sum_{\mathrm{status}_\ell = \tagdenseA}\mu_\ell\abs{X_\ell}\abs{Y_\ell} \le 2\alpha\abs{X}\abs{Y^*}$
    (\Cref{thm:2d-decomp}), summing~\eqref{eq:bucketed-rec-bound} gives
    \[
        \begin{aligned}
            \sum_{\ell : \mathrm{status}_\ell = \tagdenseA} \pot(\mathcal{S}_\ell)
            &\le \frac{7}{2}\Lambda_h 2^{-10}\alpha^{-3/4}(2\beta^*)^{1/4}\abs{Z^*} \cdot 2\alpha\abs{X}\abs{Y^*} \\
            &= \frac{7 \cdot 2^{1/4}}{1024}W^* < \frac{14}{1024}W^* < \frac{1}{64}W^*.
        \end{aligned}
    \]
    \item \emph{Regular complements ($I_{\mathrm{reg}} := \{\ell < L : \mathrm{status}_\ell = \tagreg, R_\ell \ne \emptyset\}$):}
    Here $\Delta h_\ell = 1$, $\abs{R_\ell} < \gamma\abs{Z^*}$ ($\gamma = 2^{-12}$), and
    $\dens{B[Y_\ell, R_\ell]} \le 2\beta^*$, so~\eqref{eq:bucketed-rec-bound} and the Riemann-sum bound give
    \[
        \sum_{\ell \in I_{\mathrm{reg}}} \pot(\mathcal{S}_\ell)
        \le \frac{7}{2}(2\Lambda_h)(2\beta^*)^{1/4}(\gamma\abs{Z^*}) \cdot 4(2\alpha)^{1/4}\abs{X}\abs{Y^*}
        = 28\sqrt{2}\,\gamma W^*
        < \frac{56}{4096}W^*
        < \frac{1}{64}W^*.
    \]
\end{enumerate}
The four groups total $< \frac{4}{64}W^* = \frac{1}{16}W^*$ against the $-\frac{1}{4}W^*$ credit
in~\eqref{eq:concavity-drop}, closing the induction.
\end{proof}

\subsection{Jump-Adaptive Recurrence and Proof of \texorpdfstring{\Cref{thm:main-decomp}}{Theorem 1.3}}\label{sec:3d-recurrence}

\begin{lemma}[Piece count and running time]\label{lem:recurrence}
Let $T(h)$ be the supremum of piece counts returned by valid calls at depths $h' \ge h$. Enlarging
$M_{\mathrm{cube}}, M_{\mathrm{rect}} = 2^{O(d)}$ to at least $1$, set
$Q := 2^{135 d} M_{\mathrm{cube}} M_{\mathrm{rect}} = 2^{O(d)}$ and
$\Gamma := 5d + \lceil \log_2((2d + 4)Q) \rceil = O(d)$. Then $T(H) \le 3 \cdot 2^d$ and
$T(h) \le 3 \cdot 2^{d + d^2 + \Gamma(H - h)}$ for $0 \le h < H$. Thus calling $\DecomposeThreeD$ on
$(X, Y, Z, A, B, 1, 0)$ emits $K \le T(0) \le 2^{O(d^2)}$ pieces in deterministic $O(n^{5/2} 2^{O(d^2)})$ time.
\end{lemma}

\begin{proof}
At $h = H$, a call returns $1$ or $N_{\mathrm{split}} < 3 \cdot 2^d$ pieces (\Cref{lem:potential}), so
$T(H) \le 3 \cdot 2^d$. Inducting downward on $h < H$, along a same-depth chain prior to
\Cref{line:3d-base1}, each extraction at $B$-scale $S_B = s \in \{0, \dots, d\}$ deletes
$> 2^{-d} M_{\mathrm{cube}}^{-1} 2^{-s d}\abs{Y}\abs{Z}$ edges (\Cref{lem:cube-step}), allowing at most
$K_{B,s} \le 2^d M_{\mathrm{cube}} 2^{s d}$ extractions at scale $s$ and at most
$\sum_{s=0}^{40} K_{B,s} \le 41 \cdot 2^{41 d} M_{\mathrm{cube}}$ normal extractions. Each normal
extraction yields at most $L_{\mathrm{reg}} + 1 \le 81 \cdot 2^{80 d} M_{\mathrm{rect}}$ direct pieces and
at most $d(L_{\mathrm{reg}} + 1)$ depth-$(h+1)$ children (\Cref{thm:2d-decomp}); since $41 < 2^6$,
$81 < 2^7$, and $d < 2^d$ give $3321 d < 2^{14 d}$, both chain totals are at most
$3321 d \cdot 2^{121 d} M_{\mathrm{cube}} M_{\mathrm{rect}} \le Q$, plus $1$ terminal sparse residual.
At scale $s$, at most $K_{B,s} \le Q 2^{s d}$ $B$-boosted children ($\rho_B \ge \max(2^{40}, 2^s)$) jump
by at least $\Delta_B(s) := \min(H - h, \max(20, \lfloor s/2 \rfloor))$, and at most
$(41 \cdot 2^{41 d} M_{\mathrm{cube}})(d \cdot 2^d M_{\mathrm{rect}} 2^{s d}) \le Q 2^{s d}$ $A$-boosted
children ($\mu_\ell/\alpha \ge \max(2^{40}, 2^{s-40})$) jump by at least
$\Delta_A(s) := \min(H - h, \max(20, \lfloor (s - 40)/2 \rfloor))$. Taking the supremum over valid calls
at $h' \ge h$ gives
\[
    T(h) \le 1 + Q + Q T(h+1) + \sum_{s=0}^d Q 2^{s d}\bigl(T(h + \Delta_B(s)) + T(h + \Delta_A(s))\bigr).
\]
Put $U_h := 3 \cdot 2^{d + d^2 + \Gamma(H - h - 1) + 5d}$. For $\delta \in \{\Delta_B(s), \Delta_A(s)\}$,
if $h + \delta = H$ then $2^{s d} T(H) \le 3 \cdot 2^{d + d^2} \le U_h$. If $h + \delta < H$, then
$\delta \ge \max(20, \lfloor (s - 40)/2 \rfloor)$, which implies $s < 5\delta$ ($s \le 4\delta$ for
$s \le 80$, and $s \le 2\delta + 41 < 5\delta$ for $s > 80$ as $\delta \ge 20$), so the induction
hypothesis and $\Gamma \ge 5d$ give
$2^{s d} T(h + \delta) \le 3 \cdot 2^{d + d^2 + \Gamma(H - h - \delta) + 5\delta d} \le U_h$. Also
$T(h+1) \le U_h$ and $1 + Q < Q U_h$, so $2^{\Gamma - 5d} \ge (2d + 4)Q$ yields
$T(h) \le (2d + 4)Q U_h \le 3 \cdot 2^{d + d^2 + \Gamma(H - h)}$ and $K \le T(0) \le 2^{O(d^2)}$.
Finally, every maximal chain and ceiling call appends at least one entry to $\mathcal{S}$ (even if
$B_{\mathrm{curr}} = 0$), giving $\le K = 2^{O(d^2)}$ chains, each running $\le (d + 1)2^d M_{\mathrm{cube}} 2^{d^2} = 2^{O(d^2)}$
calls to $\FindCube$ at $O(n^{5/2} 2^{O(d^2)})$ time each (\Cref{lem:cube-step}), for
$O(n^{5/2} 2^{O(d^2)})$ total deterministic time.
\end{proof}

\restatethm{thmmaindecomp}{thm:main-decomp}

\begin{proof}
Run $\DecomposeThreeD(X, Y, Z, A, B, 1, 0)$ (\Cref{alg:3d}) at even $d \ge 640$; $\max_{y \in Y}\deg_{A^T}(y) \le 1 \le 2\alpha$
holds at $\alpha = 1, h = 0$. \Cref{lem:children-invariants,lem:potential,lem:recurrence} give exactness,
piece regularity/sparsity, $\pot(\mathcal{S}) \le \Lambda_0\dens{B}^{1/4}\abs{X}\abs{Y}\abs{Z} \le 1024\,\abs{X}\abs{Y}\abs{Z}$,
$K \le 2^{O(d^2)}$, and time $O(n^{5/2} 2^{O(d^2)})$.
\end{proof}

\section{Combinatorial Triangle Detection and BMM}\label{sec:bmm}

\begin{proposition}[BMM to tripartite triangle detection~{\cite{WilliamsW18}; see also~\cite[Lemma~6.1]{AFKLM24}}]\label{prop:bmm-to-tridet}
Suppose a deterministic combinatorial algorithm solves triangle detection on $N$-vertex tripartite
graphs in worst-case time $O(N^3 / f(N))$, where $f : \N_{\ge 1} \to \R_{\ge 1}$ is non-decreasing and
$1 \le f(N) \le N^{1/2}$ for all $N \ge 1$. Then, for every integer $n \ge 1$, there is a deterministic
combinatorial algorithm computing the Boolean product $A \star B$ of two $n \times n$ Boolean matrices
in worst-case time $O(n^3 / f(\lfloor n^{1/3}\rfloor))$.
\end{proposition}

\restatethm{thmmainbmm}{thm:main-bmm}

\begin{proof}
On a tripartite instance $(X, Y, Z; A, B, C)$ ($C \in \{0,1\}^{X \times Z}$, $n := \abs{X} + \abs{Y} + \abs{Z}$),
return \textnormal{\textsc{False}} in $O(1)$ time if any part is empty; otherwise $n \ge 3$. Let
$C_{\mathrm{dec}} \ge 1$ bound both $2^{O(d^2)}$ exponents in \Cref{thm:main-decomp}, fix $\varepsilon := 1/160$,
and set $c_0 := 1/\lceil 640\sqrt{C_{\mathrm{dec}}} \rceil$,
$t := \lfloor c_0\sqrt{\lfloor \log_2 n \rfloor}\rfloor$, and $d := 320\max(2, t)$, so even $d \ge 640$
gives an exact integer $q := \varepsilon d/2 = d/320 = \max(2, t) \ge 2$ with $2^d = O(n^{1/2})$ and
$2^{C_{\mathrm{dec}} d^2} = O(n^{1/4})$ (\Cref{app:applications}). By \Cref{thm:main-decomp}, decomposing
$(A, B)$ takes $O(n^{5/2} 2^{C_{\mathrm{dec}} d^2}) = O(n^{11/4})$ time and yields
$K \le 2^{C_{\mathrm{dec}} d^2} = O(n^{1/4})$ pieces with $AB = \sum_{k=1}^K A_k B_k$ and
$\sum_{k=1}^K w_k = \pot(\mathcal{S}) \le 1024\,n^3$ ($w_k := \dens{A_k}^{1/4}\dens{B_k}^{1/4}\abs{X_k}\abs{Y_k}\abs{Z_k}$).
Building supports in $O(K n^2) = O(n^{9/4})$ total time, we test each piece $k \in [K]$ against
$C_k := C[X_k, Z_k]$ (writing $u := \dens{A_k}$, $v := \dens{B_k}$, $m_k := \min(u, v, \dens{C_k})$):
\begin{enumerate}[nosep,leftmargin=*,topsep=2pt]
    \item \emph{If $\abs{Y_k}\abs{Z_k} < 2^d$:} Scan $X_k \times Y_k \times Z_k$ in $O(n 2^d) = O(n^{3/2})$
    time ($O(n^{7/4})$ total across all pieces).
    \item \emph{Else if $m_k \le 2^{-q} = 2^{-\varepsilon d / 2}$:} Scan the support of a minimum-density
    matrix among $A_k, B_k, C_k$ across the opposite vertex set. Whether $\min(u, v) \le 2^{-\varepsilon d / 2}$
    (so $m_k \le \min(u, v) \le 2^{-\varepsilon d / 4} u^{1/4} v^{1/4}$) or $\min(u, v) > 2^{-\varepsilon d / 2}$
    (so $1 < 2^{\varepsilon d / 4} u^{1/4} v^{1/4}$ and $m_k \le 2^{-\varepsilon d / 2} < 2^{-\varepsilon d / 4} u^{1/4} v^{1/4}$),
    this checks $m_k\abs{X_k}\abs{Y_k}\abs{Z_k} \le 2^{-\varepsilon d / 4} w_k = 2^{-d/640} w_k$ triples,
    summing to $O(K + 2^{-d/640}\pot(\mathcal{S})) = O(n^3 / 2^{d/640}) = O(n^3 / 2^{\Omega(\sqrt{\log n})})$.
    \item \emph{Else ($u, v, \dens{C_k} > 2^{-\varepsilon d / 2} > 2^{-d}$ and $\abs{Y_k}\abs{Z_k} \ge 2^d$):}
    Piece $k$ is regular by \Cref{thm:main-decomp}, so \Cref{prop:regular-product} ensures at most a
    $2^{-\varepsilon d / 2}$-fraction of $X_k \times Z_k$ has $(A_k B_k)(x, z) = 0$; since
    $\dens{C_k} > 2^{-\varepsilon d / 2}$, the pigeonhole principle certifies a triangle immediately---return
    \textnormal{\textsc{True}}.
\end{enumerate}
Returning \textnormal{\textsc{False}} if no piece finds a triangle solves $N$-vertex tripartite triangle
detection in $O(N^3 / 2^{\Omega(\sqrt{\log N})})$ time. The three-copy reduction for general graphs and
\Cref{prop:bmm-to-tridet} with $f(N) := \min(N^{1/2}, 2^{c_1\sqrt{\log_2 N}})$ complete \Cref{thm:main-bmm}
(\Cref{app:applications}).
\end{proof}

\section{AI Disclosure}
\label{sec:ai-disclosure}
This work was carried out with substantial assistance from Gemini~4 agents. The authors specified the research topic, the target problem, and the high-level direction. Gemini~4 agents developed the proofs and algorithmic constructions in all phases of the project. The Gemini~4 agents also drafted and revised the manuscript and produced the Lean~4 formalization. The human authors verified and further edited the paper for clarity of presentation, and take full responsibility for the final version.

Formal versions of \Cref{thm:main-bmm,thm:main-decomp} and \Cref{prop:regular-product} are
machine-checked in Lean~4 (zero \texttt{sorry} declarations, Lean's three standard axioms only).
The formalization follows the full version of this paper, available in the
\href{https://github.com/lockqueue/bmm}{accompanying repository}, whose algorithm uses a
two-stage sampler and different constants from those printed here; the formal statements imply
the bounds of \Cref{thm:main-bmm,thm:main-decomp}. The
cited sampler and extractor bounds~\cite{GoldreichW97,AFKLM24,GUV09,Vadhan12} enter as explicit
hypotheses, matrices are represented noncomputably over $\R$, running time is tracked via an
explicit cost-accounting monad (\texttt{CostM}) rather than a Word-RAM model, and the black-box
applications of \Cref{rem:applications} are not formalized. Full details appear in
\texttt{lean/FORMALIZATION.pdf} in the repository.

\clearpage
\appendix
\crefalias{section}{appendix}

\section{Deferred Sampling and Grid-Norm Runtime Proofs}\label{app:sampling}

\begin{proof}[Proof of \Cref{prop:sampler}]
Set $x := 2\abs{Y}/\varepsilon_{\mathrm{samp}} > 2$, and let $N = 2^\ell \ge 4$ be the smallest power of
two at least $\lceil x \rceil$; since $\lceil x \rceil < x + 1 < \frac{3}{2}x$, we have
$2\abs{Y}/\varepsilon_{\mathrm{samp}} \le N < 2\lceil x \rceil < 3x = 6\abs{Y}/\varepsilon_{\mathrm{samp}}$.
Write $N = q_0\abs{Y} + r_0$ with $q_0 := \lfloor N/\abs{Y}\rfloor$ and $0 \le r_0 < \abs{Y}$, and define
$\pi : \{0, 1\}^\ell \to Y$ by repeating each element of $Y$ $q_0$ times and filling the remaining $r_0$
slots arbitrarily; then every $g : Y \to [0, 1]$ obeys
$\abs{\E_{u \in \{0, 1\}^\ell}[g(\pi(u))] - \E_{y \in Y}[g(y)]} \le r_0/N < \abs{Y}/N \le \varepsilon_{\mathrm{samp}}/2$.
By~\cite[Corollary~6.24 and Theorem~6.36]{Vadhan12} (combining~\cite[Proposition~2.7]{Zuckerman97}
and~\cite[Theorem~1.5]{GUV09} at min-entropy $k := \lceil \frac{5}{4}\ell \rceil$, source length
$m := k + \lceil \log_2(4/\delta_{\mathrm{samp}})\rceil$, and error $\varepsilon_{\mathrm{samp}}/4$,
truncating its $\ge \frac{4}{5}k \ge \ell$ output bits to $\ell$), one constructs in deterministic time
$2^m\poly(m, \varepsilon_{\mathrm{samp}}^{-1}) = \abs{Y}^{5/4}\poly(\log(\abs{Y} + 2), \varepsilon_{\mathrm{samp}}^{-1}, \delta_{\mathrm{samp}}^{-1})$
an explicit $(\varepsilon_{\mathrm{samp}}/2, \delta_{\mathrm{samp}})$-oblivious sampler $\mathcal{T}_0$ on
$\{0, 1\}^\ell$ for $[0, 1]$-valued functions with two-sided error, with
$\abs{\mathcal{T}_0} = 2^m < 16 N^{5/4}\delta_{\mathrm{samp}}^{-1} = O(\abs{Y}^{5/4}\varepsilon_{\mathrm{samp}}^{-5/4}\delta_{\mathrm{samp}}^{-1})$
and $\abs{T_0} \le \poly(m, \varepsilon_{\mathrm{samp}}^{-1})$. Pushing each sequence forward via
$\mathcal{T} := (\pi(T_0))_{T_0 \in \mathcal{T}_0}$ and applying $\mathcal{T}_0$ to $g \circ \pi$
gives~\eqref{eq:oblivious-sampler} by the triangle inequality.
\end{proof}

\begin{remark}[Choice of oblivious sampler]\label{rem:sampler-choice}
Prior work~\cite[Lemma~4.4]{AFKLM24} used the Goldreich--Wigderson sampler~\cite{GoldreichW97}, whose
sample length $\abs{T}$ is $\poly(\delta_{\mathrm{samp}}^{-1})$. In $\EstimateNorm$ (\Cref{sec:2d-norm}),
we invoke the sampler at $\delta_{\mathrm{samp}} = (\alpha'/2)^d = 2^{-\Theta(d^2)}$ and enumerate
$d$-tuples in $T^d$, so $\abs{T} = \poly(\delta_{\mathrm{samp}}^{-1})$ would inflate $\abs{T}^d$ to
$2^{\Theta(d^3)}$. By contrast, \Cref{prop:sampler} keeps $\abs{T} = \poly(\log n, 2^{O(d)})$ at the mild
cost of family size $\abs{\mathcal{T}} \propto \abs{Y}^{5/4}$ instead of $\abs{Y}$. While two-stage
composition~\cite{GoldreichW97,Goldreich11,GUV09} would recover $O(n^2 2^{O(d^2)})$ decomposition time, our
single-stage bound $O(n^{5/2} 2^{O(d^2)})$ already evaluates to $O(n^{11/4}) \ll n^3 / 2^{\Omega(\sqrt{\log n})}$
at $d = \Theta(\sqrt{\log n})$ in \Cref{sec:bmm}, avoiding sampler composition at zero cost to \Cref{thm:main-bmm}.
\end{remark}

\begin{proof}[Running-time and bit-length proof for \Cref{lem:approx-grid-norm}]
By \Cref{prop:sampler}, Step~1 constructs $\mathcal{T}$ in time
$O(n^{5/4}(\log_2(n + 2))^{O(1)}(2d/\alpha')^{O(d)})$ with $\abs{\mathcal{T}} = O(n^{5/4}(2/\alpha')^{d+2})$
and $t := \abs{T} \le (\log_2(n + 2))^C(2d/\alpha')^C$ for a constant $C \ge 1$. Set
$u := \log_2(n + 2) \ge 2$; applying the Taylor-series inequality $x < 1 + x < e^x < 4^x = 2^{2x}$ at
$x := \frac{u}{16 C d} > 0$ gives $\frac{u}{16 C d} < 2^{u / (8 C d)}$, and raising to the $C d$-th power
yields
\[
    \bigl(\log_2(n + 2)\bigr)^{C d}
    = u^{C d}
    < (16 C d)^{C d} 2^{u/8}
    = (16 C d)^{C d}(n + 2)^{1/8}
    = O\!\left(n^{1/8}(d/\alpha')^{O(d)}\right),
\]
since $(16 C)^C = O(1)$ at $d = 1$ and $d/\alpha' > 2$ absorbs $2^{O(d)}$ for $d \ge 2$. Thus
$\abs{\mathcal{T}} t^d = O(n^{11/8}(d/\alpha')^{O(d)})$. Maintaining common denominators keeps all
rationals in Steps~2--3 within $O(d(\log_2(n + 2) + \log_2(2d/\alpha')))$ bits, so each arithmetic update
takes $\poly(\log_2(n + 2), d, \log_2(2/\alpha')) = O(n^{1/8}(d/\alpha')^{O(d)})$ word operations; two
passes over $X$ ($\abs{X} \le n$) per tuple therefore take
$O(n \cdot n^{11/8} \cdot n^{1/8}(d/\alpha')^{O(d)}) = O(n^{5/2}(d/\alpha')^{O(d)})$ total time.
\end{proof}

\section{Deferred Proofs for Decomposition Helpers and Invariants}\label{app:helper-proofs}

\begin{proof}[Proof of \Cref{lem:riemann}]
If $L = 1$, $A_1 = A$ and $1 < 1/p$. For $L \ge 2$, set $t_\ell := \dens{A^{(\ell)}}$. Every
$\ell \in [L-1]$ satisfies $\dens{A_\ell} \ge t_\ell > 0$ (\Cref{lem:find-rect}) and
$t_\ell - t_{\ell+1} = \frac{\abs{X_\ell}\abs{Y_\ell}}{\abs{X}\abs{Y}}\dens{A_\ell}$. Since
$t \mapsto t^{-(1-p)}$ is decreasing on $(0, \infty)$ and integrable at $0$ (as $p > 0$),
\[
    \frac{\abs{X_\ell}\abs{Y_\ell}}{\abs{X}\abs{Y}}\dens{A_\ell}^p
    = \dens{A_\ell}^{-(1-p)}(t_\ell - t_{\ell+1})
    \le t_\ell^{-(1-p)}(t_\ell - t_{\ell+1})
    \le \int_{t_{\ell+1}}^{t_\ell} t^{-(1-p)}\,dt
    = \frac{1}{p}\bigl(t_\ell^p - t_{\ell+1}^p\bigr).
\]
Summing over $\ell \in [L-1]$ and adding $t_L^p \le \frac{1}{p}t_L^p$ telescopes to
$\frac{1}{p}t_1^p = \frac{1}{p}\dens{A}^p$, proving~\eqref{eq:riemann-telescoping}.
\end{proof}

\begin{proof}[Proof of \Cref{lem:dyadic-bucketing}]
Since $2^d\mu > 1 \ge \deg_{(A')^T}(y)$, the buckets~\eqref{eq:dyadic-buckets} partition $Y'$.
Restricting $Y'$ leaves column averages over $X'$ unchanged, giving degree cap $2^{j+1}\mu = 2\alpha'_j$,
and $\alpha'_0 = \mu \le 1$ while any $y \in Y'_j$ ($j \ge 1$) has $2^j\mu < \deg_{(A')^T}(y) \le 1$. On
non-empty buckets, put $p_j := \abs{Y'_j}/\abs{Y'}$ and $\beta'_j := \dens{B'[Y'_j, Z']}$. Markov's
inequality on $\E_{y \in Y'}[\deg_{(A')^T}(y)] \le \mu$ gives $p_j \le 2^{-j}$ for all $j \ge 0$, while
$\sum_j p_j\beta'_j = \dens{B'}$ gives $p_j\beta'_j \le \dens{B'}$. Summing over non-empty buckets yields
\[
    \sum_j (\alpha'_j)^{1/4}(\beta'_j)^{1/4}\abs{Y'_j}
    = \abs{Y'}\sum_j (2^j\mu)^{1/4} p_j^{3/4}(p_j\beta'_j)^{1/4}
    \le \mu^{1/4}\dens{B'}^{1/4}\abs{Y'}\sum_{j=0}^{d-1} 2^{-j/2},
\]
and $\sum_{j=0}^{d-1} 2^{-j/2} < (1 - 2^{-1/2})^{-1} = 2 + \sqrt{2} < 7/2$.
\end{proof}

\begin{proof}[Proof of \Cref{lem:cube-step}]
In a round entered at $b := \dens{B} \in (2^{-d}, 2^{40}\beta_{\mathrm{init}})$, row pruning
(\Cref{line:cube-prune1,line:cube-rec1}) retains $\abs{Y'} \ge \frac{1}{3}\abs{Y}$ (\Cref{lem:min-degree})
and either restarts or certifies $B'$ as $(\varepsilon\gamma/2)$-min-degree with $b' := \dens{B'} \ge b$,
so every non-empty $Y_\ell \subseteq Y'$ has $\dens{B[Y_\ell, Z]} \ge (1 - \frac{\varepsilon\gamma}{2})b'$.
Since $[2^{-40}\alpha, 2^{40}\alpha]$ has ratio $2^{80}$, \Cref{lem:find-rect,thm:2d-decomp} bound each
search in $\DecomposeTwoD$ (\Cref{line:cube-dec2d}) to $r_{\mathrm{rect}} \le 1 + 160/\varepsilon$ steps,
retaining $\abs{Y_\ell} \ge 2^{-c_1 d}\abs{Y'} \ge \frac{1}{3} 2^{-c_1 d}\abs{Y}$ for
$c_1 := 2(1 + 160/\varepsilon)$ and all $\ell \in [L]$. For $\ell \in [L-1]$, the slice check
(\Cref{line:cube-slice}) either restarts at density $\ge 2b' \ge 2b$ or leaves $\dens{B[Y_\ell, Z]} < 2b'$,
covering all deferred $A$-boosted slices ($\mathrm{status}_\ell = \tagdenseA$). On a regular slice
($\mathrm{status}_\ell = \tagreg$), column pruning (\Cref{line:cube-prune2,line:cube-rec2}) retains
$\abs{Z_\ell} \ge \frac{1}{3}\abs{Z} > 0$; a restart achieves
$\dens{B_\ell} \ge (1 + \varepsilon\gamma)(1 - \frac{\varepsilon\gamma}{2})b \ge (1 + \frac{\varepsilon\gamma}{4})b$
($\varepsilon\gamma < 1/2$), and otherwise $B_\ell^T$ is $\varepsilon$-min-degree with
$\abs{Z \setminus Z_\ell} < \gamma\abs{Z}$, $(1 - \frac{\varepsilon\gamma}{2})b' \le \dens{B_\ell} < 2(1 + \varepsilon\gamma)b' < 3b'$,
and $\dens{B[Y_\ell, Z \setminus Z_\ell]} < (1 - \varepsilon)\dens{B_\ell} < 2b'$ when
$Z \setminus Z_\ell \ne \emptyset$ (as $(1 - \varepsilon)(1 + \varepsilon\gamma) < 1$). Since
$(1 + \frac{\varepsilon}{2})(1 - \frac{\varepsilon\gamma}{2}) > 1 + \frac{\varepsilon\gamma}{4}$, we have
$(1 + \frac{\varepsilon}{2})\dens{B_\ell} > 2^{-d}$, so \Cref{lem:dyadic-sift} applies to $B_\ell^T$
(\Cref{line:cube-sift}): a witness $(Z'', Y'', s)$ gives
$\dens{B[Y'', Z'']} \ge 2^s(1 + \frac{\varepsilon\gamma}{4})b$ and
$\abs{Y''}\abs{Z''} \ge \frac{\varepsilon}{16} 2^{-(s+4)d}\abs{Y_\ell}\abs{Z_\ell} \ge \frac{\varepsilon}{144} 2^{-(s + 4 + c_1)d}\abs{Y}\abs{Z}$,
and otherwise $B_\ell^T$ is $(\varepsilon, 2, d)$-regular.

With $a := \varepsilon\gamma/4$ and $\eta := \frac{\varepsilon}{144} 2^{-(4+c_1)d}$, each restart
multiplies density by $\ge 2^{s_t}(1 + a)$ and area by $\ge \eta 2^{-s_t d}$. If $r \ge 1$, round
$r - 1$ enters below $2^{40}\beta_{\mathrm{init}}$, so $2^{(r-1)a} \le (1 + a)^{r-1} < 2^{40}$ ($2^x \le 1 + x$
on $[0, 1]$), giving $r \le R_{\mathrm{cube}}$. Multiplying across rounds gives
$1 \ge \beta^* \ge 2^{S_B}\beta_{\mathrm{init}} > 2^{S_B - d}$ ($S_B \le d$) and
$\abs{Y^*}\abs{Z^*} \ge \frac{1}{3}\eta^r 2^{-S_B d}\abs{Y}\abs{Z} \ge M_{\mathrm{cube}}^{-1} 2^{-S_B d}\abs{Y}\abs{Z}$
for $M_{\mathrm{cube}} := 3\eta^{-R_{\mathrm{cube}}} = 2^{O(d)}$. A normal return passes the cap and
row-restart tests, so $\beta^* < 2^{40}(1 + a)\beta_{\mathrm{init}} < 2^{41}\beta_{\mathrm{init}}$ and
$S_B \le 40$, while \Cref{thm:2d-decomp} gives the $A_\ell$ bounds. Over $r + 1 = O(1)$ rounds, row
pruning, $\DecomposeTwoD$, $L \le 2^{O(d^2)}$ slice checks ($O(n^2)$ each), and $L_{\mathrm{reg}} = 2^{O(d)}$
column-pruning and $\Sift$ calls take $O(n^{5/2} 2^{O(d^2)})$ time.
\end{proof}

\begin{proof}[Proof of \Cref{lem:children-invariants}]
In an interior call, $h < H$, $\dens{A}, \dens{B} > 2^{-d}$, and $\abs{Y}\abs{Z} \ge 2^d$. By
\Cref{lem:cube-step}, $Y^*, Z^*$ are non-empty with $\dens{B[Y^*, Z^*]} \ge \dens{B} > 0$, so the
residual child (\Cref{line:3d-sstar}) strictly reduces $\nnz(B - B[Y^*, Z^*]) < \nnz(B)$ at depth $h$.
The $B$-boosted child (\Cref{line:3d-recb}) keeps $X, \alpha$ fixed (preserving column degrees on
$Y^* \subseteq Y$) with $\rho_B \ge 2^{40}$, so $\Delta h_B \in [1, H - h]$. Each bucketed child
(\Cref{line:3d-reca}) has non-empty $X_\ell, Y_{\ell,j}, R_\ell$, $\Delta h_\ell \in [1, H - h]$
($\mu_\ell/\alpha \ge 2^{40}$ on $A$-boosted slices gives jump $\ge 20$, and $\Delta h_\ell = 1$
otherwise), and $\mu_\ell \in (2^{-d}, 1]$: by \Cref{lem:cube-step}(1) on regular slices, by
$\mu_\ell \ge 2^{40}\alpha > 2^{39 - d} > 2^{-d}$ on $A$-boosted slices ($2^{-d} < \dens{A} \le 2\alpha$),
and by \Cref{line:cube-zl} for $\ell = L$ ($Z_L = \emptyset$). Thus \Cref{lem:dyadic-bucketing} gives
$\alpha'_j = 2^j\mu_\ell \in (0, 1]$ and column cap $2\alpha'_j$, proving~(1) and termination.

We prove~(2) and~(3) by induction on $(H - h, \nnz(B))$. The sparse base case (\Cref{line:3d-base1})
returns one sparse piece. At $h = H$ (\Cref{line:3d-base2}), $\abs{Y}\abs{Z} \ge 2^d$ ensures
$m = \lfloor 2^{-d}\abs{Y}\abs{Z} \rfloor \ge 1$; the batches partition $\supp(B)$ with
$AB = \sum_{s=1}^{N_{\mathrm{split}}} A B_s$, and every batch has $\dens{B_s} \le m/(\abs{Y}\abs{Z}) \le 2^{-d}$
(sparse). For $h < H$, writing $B^* := B[Y^*, Z^*]$, bilinearity gives $AB = A(B - B^*) + A[X, Y^*]B^*$.
Induction handles $A(B - B^*)$ via $\mathcal{S}^*$ and, in the $B$-boosted branch, $A[X, Y^*]B^*$ too;
in the normal branch, $A[X, Y^*] = \sum_{\ell=1}^L A_\ell$ and $Y_\ell = \bigsqcup_{j=0}^{d-1} Y_{\ell,j}$
expand $A[X, Y^*]B^*$ into
\[
    \sum_{\ell \in [L] : Z_\ell \ne \emptyset} A_\ell B[Y_\ell, Z_\ell]
    + \sum_{\ell \in [L] : R_\ell \ne \emptyset}
      \sum_{j : Y_{\ell,j} \ne \emptyset} A_\ell[X_\ell, Y_{\ell,j}] B[Y_{\ell,j}, R_\ell],
\]
matching $\mathcal{S}_{\mathrm{direct}}$ plus $\bigcup_\ell \mathcal{S}_\ell$ with entrywise dominance.
Finally, in $\mathcal{S}_{\mathrm{direct}}$, a direct remainder ($\ell = L$) has $\dens{A_L} \le 2^{-d}$
(sparse), while each $\ell < L$ with $Z_\ell \ne \emptyset$ is a regular slice ($\mathrm{status}_\ell = \tagreg$):
sparse if $\dens{B_\ell} \le 2^{-d}$ or $\abs{Y_\ell}\abs{Z_\ell} < 2^d$, and regular otherwise by
\Cref{lem:cube-step}(1).
\end{proof}

\section{Word-RAM Implementation Details and Combinatorial Applications}\label{app:applications}

\paragraph{Exact integer arithmetic and general-to-tripartite reduction in \Cref{thm:main-bmm}.}
In the proof of \Cref{thm:main-bmm}, let $b_0 := \lceil 640\sqrt{C_{\mathrm{dec}}} \rceil \ge 640$ (an
explicit positive integer constant) and $c_0 := 1/b_0$. Given $n \ge 3$, we compute $L_0 := \lfloor \log_2 n \rfloor$
by repeated doubling, $t := \lfloor \sqrt{L_0}/b_0 \rfloor$ as the unique integer with
$b_0^2 t^2 \le L_0 < b_0^2(t + 1)^2$, $q := \max(2, t)$, $d := 320 q$, and the exact powers of two $2^q$
and $2^d$ in $O(\log n)$ elementary integer operations without floating-point arithmetic, testing
$m_k \le 2^{-q}$ by exact integer multiplication by $2^q$. Since $320 c_0 \le 1/2$ and
$C_{\mathrm{dec}}(320 c_0)^2 \le 1/4$, for $t \ge 2$ we have $d = 320 t \le \frac{1}{2}\log_2 n$ and
$C_{\mathrm{dec}} d^2 \le \frac{1}{4}\log_2 n$ (while $d = 640 = O(1)$ for $t < 2$), giving
$2^d = O(n^{1/2})$ and $2^{C_{\mathrm{dec}} d^2} = O(n^{1/4})$. Finally, for triangle detection on a
general $n$-vertex undirected graph $G = (V, E)$, returning \textnormal{\textsc{False}} when $n < 3$ and
otherwise forming three disjoint copies $(V_1, V_2, V_3)$ of $V$ with edges $(u_i, v_j)$ whenever
$\{u, v\} \in E$ ($i < j$) reduces in $O(n^2)$ time to a $3n$-vertex tripartite instance.

\begin{remark}[Combinatorial applications]\label{rem:applications}
Plugging \Cref{thm:main-bmm} into known black-box combinatorial reductions yields matching deterministic
combinatorial speedups for:
\begin{itemize}[nosep,leftmargin=*,topsep=2pt]
    \item \emph{$k$-Clique} in $O(n^k / 2^{\Omega(\sqrt{\log n})})$ time for every fixed $k \ge 3$~\cite{NesetrilP85,AbboudFS24,AFKLM24};
    \item \emph{$k$-Dominating Set} in $O(n^{k+1} / 2^{\Omega(\sqrt{\log n})})$ time for every fixed $k \ge 3$~\cite{EisenbrandG04}; and
    \item \emph{All-Edges Triangle}~\cite{WilliamsW18}, \emph{context-free grammar parsing}~\cite{Valiant75,Lee02},
    \emph{transitive closure} of directed graphs~\cite{FischerM71,Munro71}, and \emph{Boolean binary join-project
    queries}~\cite{AmossenP09,GuchtWWZ15}, each in $O(n^3 / 2^{\Omega(\sqrt{\log n})})$ time.
\end{itemize}
\end{remark}

\end{document}